\documentclass{jfm}
\usepackage{graphicx}
\usepackage[figuresright]{rotating}
\usepackage{natbib}
\usepackage{epstopdf}
\usepackage{amsmath}
\usepackage{amssymb}
\usepackage{xcolor}

\ifCUPmtlplainloaded \else
  \checkfont{msam10}
  \iffontfound
    \IfFileExists{amssymb.sty}
      {\typeout{^^JFound AMS Symbol fonts on the system, using the
                'amssymb' package.^^J}%
       \usepackage{amssymb}%
         
       \let\ge=\geqslant  
      }{}
  \fi
\fi

\ifCUPmtlplainloaded \else
  \IfFileExists{amsbsy.sty}
    {\typeout{^^JFound the 'amsbsy' package on the system, using it.^^J}%
     \usepackage{amsbsy}}
    {}
\fi
\usepackage{graphicx}
\usepackage[figuresright]{rotating}
\usepackage{epstopdf}

\title{{The motion of intruders through soft solids}}

\author[T. Mullin, T. Pettinari \& J.A. Dijksman]{ \ns T. Mullin$^1$\thanks{Email address for correspondence: tom.mullin@maths.ox.ac.uk}, \ns T. Pettinari$^2$ \& \ns J.A. Dijksman$^2$}
  
\affiliation{
$^1$Mathematical Institute,  University of
Oxford OX2 6GG, United Kingdom. \\$^2$Institute of Physics, Van der Waals-Zeeman Institute, University of Amsterdam, PO Box 94216, 1090 GE Amsterdam, The Netherlands}

\makeatletter
\gdef\@underjournal{}
\def\@maketitle#1{%
 \newpage
 \vspace*{10\p@}%
 {\centering \sloppy
  {\normalfont\LARGE\fontswitch\bfseries \@title \par}%
  \vskip 14\p@ \@plus 2\p@ \@minus 1\p@
  {\normalfont\large\fontswitch\bfseries\baselineskip=12\p@
     \lowercase{\@author}\par}%
  \vskip 4\p@ \@plus 1\p@
  {\normalfont\small \@affiliation \par}%
 \par}%
 \vskip 8\p@ \@plus 2\p@ \@minus 1\p@
}
\makeatother

\begin{document}

\maketitle
\begin{abstract}
We present the results of an experimental investigation into the motion of large buoyant rigid spheres rising through highly concentrated collections of hydrated hydrogel particles. The concentration of the soft highly poro-elastic particles is such that the mechanical properties of the material are that of a soft solid which can also flow very slowly. Despite the established time-dependent, non-Newtonian character of hydrogel packings, we find that when the upper surface of the material is free, an immersed buoyant sphere travels through the material at a constant speed. Qualitatively distinct behavior is found when a rigid lid is placed on the surface of the material. In these cases, sublinear time-dependent motion of the sphere is found. The effects of the motion are generally observed to be highly localised around the sphere in all cases. However, when the translational speed of the sphere is constant, it is accompanied by significant flow at the surface of the sample whereas surface movement is suppressed when a lid is present. When  the stress exerted on the material is changed by varying the mass of the sphere, its terminal velocity is found to depend exponentially on buoyancy. We use these observations to support a hypothesis which links the exponential stress dependence of the drag coefficient induced by the material to the effects of the boundary conditions on the kinematics of the intruder.

\end{abstract}


\section{Introduction}

We report the results of an experimental investigation into the slow motion of large rigid spheres through a highly deformable soft solid. The soft material comprises a highly concentrated collection of fully hydrated hydrogels immersed in water. The particle concentration is such that the swollen particles form a dense packing  in the form of an interconnected network of highly elastic, porous material with no detectable connected regions of water in the interior or at the surface of the sample. The water maintains the hydrogels at an equilibrium swelling condition, hence attractive forces between fully swollen hydrogel spheres are minimal: the chemical potential difference~\citep{yoon2010poroelastic} between neighbouring, fully swollen particles, which may induce a negative solvent pressure, is assumed to be small. The material response of such packings is known to comprise of both  fluid-like rate dependence and solid-like elasto-plastic resistance. Hence, an embedded intruder which moves through the material will experience a drag force as a result of  elasto-plastic deformation. Moreover, the material is compressed and expands smoothly as the intruder passes through the packing. Applications of such materials are wide-ranging and can  be found in soft robotics and soil sciences; see for example~\citep{tabuteau}. 

In the laboratory, submersed hydrogel particle packings  are contained within well-defined boundaries. The  two-phase material comprises water and hydrogel spheres. Each component responds differently to end boundaries: specifically, no-slip applies to the water along rigid walls but the particles slip with respect to any surface. In the sample creation process, the dry  particles are added to water and they swell and  settle into a submersed bed. We have shown that the particle packing boundary has a significant effect on the mechanical properties of the medium. Specifically,  an external pressure exerted on the particle boundary exponentially affects the drag forces of an intruder  moving through the sample~\citep{2022creepcontrol, 2024creepconf}. Here we now show that the presence of a free surface has a dramatic effect on the drag force. While a rigid end imposed on the boundary orthogonal to the imposed stress induces a time-dependence in the speed of an intruder driven at constant stress, a free boundary condition allows an intruder under constant stress to attain a constant ``terminal'' velocity. As previously reported by \citep{kudrollihydrogel} the flow field around a moving intruder inside a hydrogel packing is highly localized at the sides of the moving intruder. However, here we find that the deformation field in front of the intruder affects the end boundary shape significantly when it is free. 

Intruders moving under constant stress conditions have been used in the past to provide  insights into the flow behavior of non-Newtonian fluids, such as viscoelastic materials~\citep{nNexp,Tabuteau2007,Sgreva}. For example, in a study of a shear thinning fluid it is found that the distance the sphere travels is $\propto t^{1/2}$ \citep{Mckinley}. An explanation for this non-linear behaviour is provided by \citep{beris1985creeping} who postulate that a build up of stagnant fluid around the sphere will lead to an increase in its effective size over time and hence the drag force increases accordingly. In shear thickening fluids, a similar coupling has been observed between the effective size of the intruder and the viscous response of the material surrounding it for the non-linear slowdown of an intruder~\citep{brassard2021viscous}, which may lead to oscillatory velocity fluctuations~\citep{2013Kann}. These examples highlight that for non-Newtonian fluids, the observed intruder kinematics can only be reliably interpreted when a detailed knowledge of the constitutive relationship of the fluid. Our work adds the important conclusion that the role of the deformation constraints placed by the end boundary conditions on the kinematics of an intruder in a fluid-like material must also be considered. The results indicate the tantalizing possibility that boundary variations can be used as a rheometric tool to elucidate the flow behavior of non-Newtonian fluids.

The focus of our current study is on a class of non-Newtonian materials, namely packings of hydrated hydrogel particles. These packings have many remarkable features, including sensitivity to the frictional properties of the hydrogels, non-Newtonian flow profiles, critical packing fraction behavior in their high flow rate rheology and various forms of creep behavior~\citep{workamp2019contact, PhysRevFluids.3.084303, gueslguazzelli, 2022creepcontrol, 2024creepconf}. Control of the properties of hydrogel suspensions in water can be achieved via the friction of hydrogel particles~\citep{workamp2019contact}, packing density~\citep{shewan2015viscosity, shewan2021viscoelasticity, gueslguazzelli, 2022creepcontrol} and even using the valency of added ions~\citep{budtova1994rheological}. In previous studies, the non-Newtonian fluid properties of dense hydrogel suspensions is known to be paramount. For example, hydrogel suspensions have been shown to behave as Herschel-Bulkley fluids~\citep{workamp2019contact} when probed with standard rheological methods under steady state, constant volume and constant driving rate conditions.  

\begin{figure}
\centering
{\includegraphics[width=4.3cm]{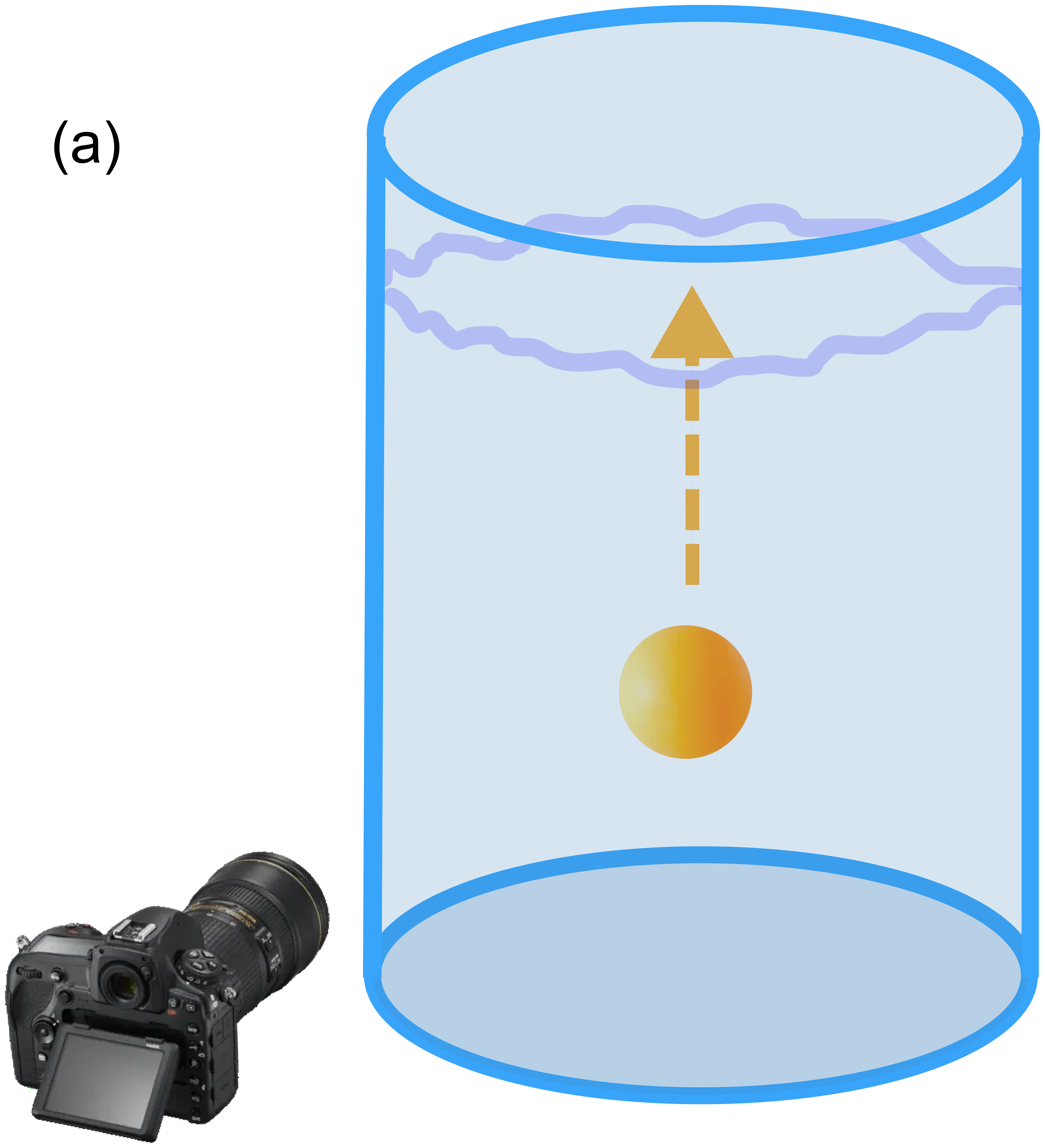}}
{\includegraphics[width=3.3cm]{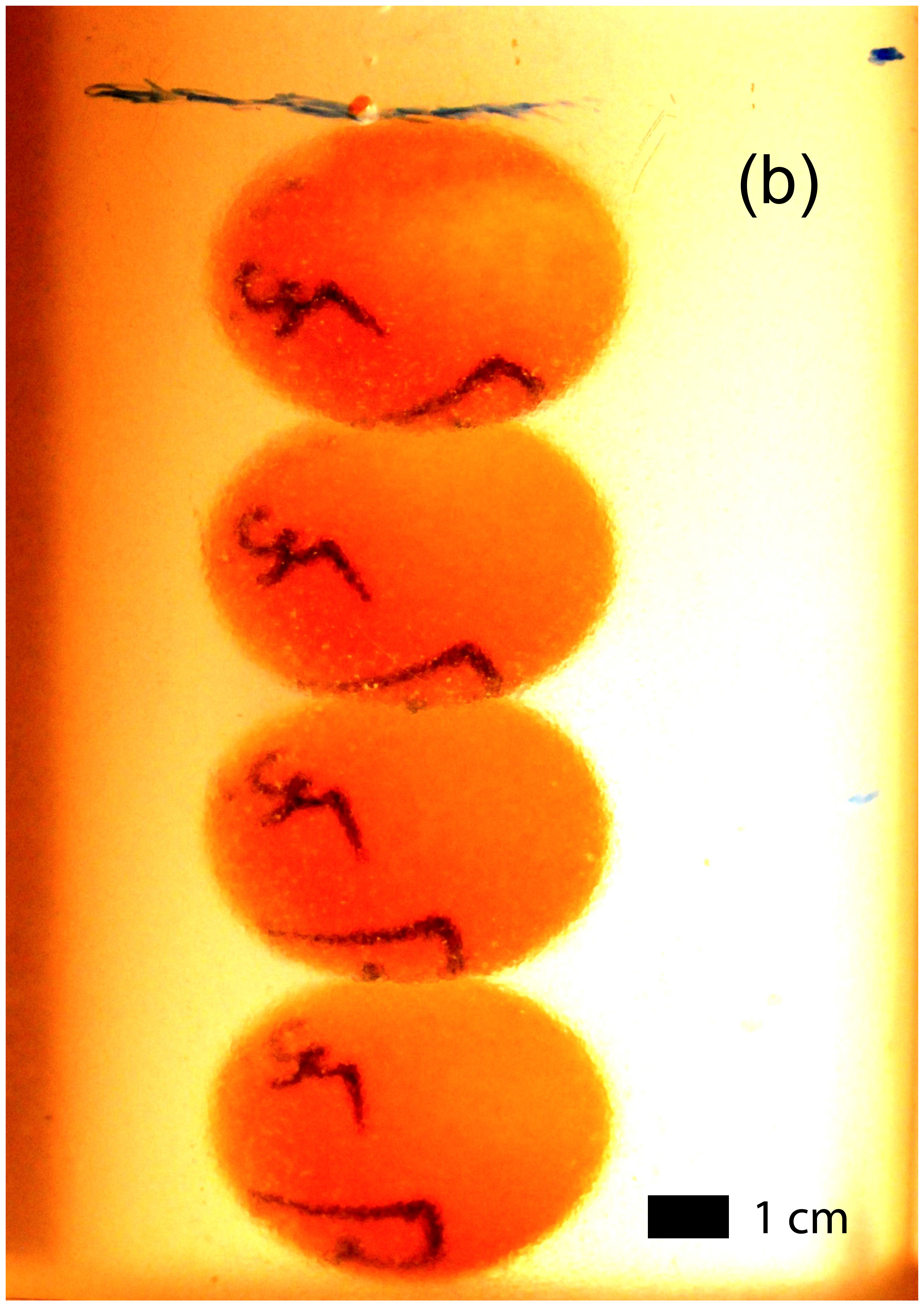}}
{\includegraphics[width=3.6cm]{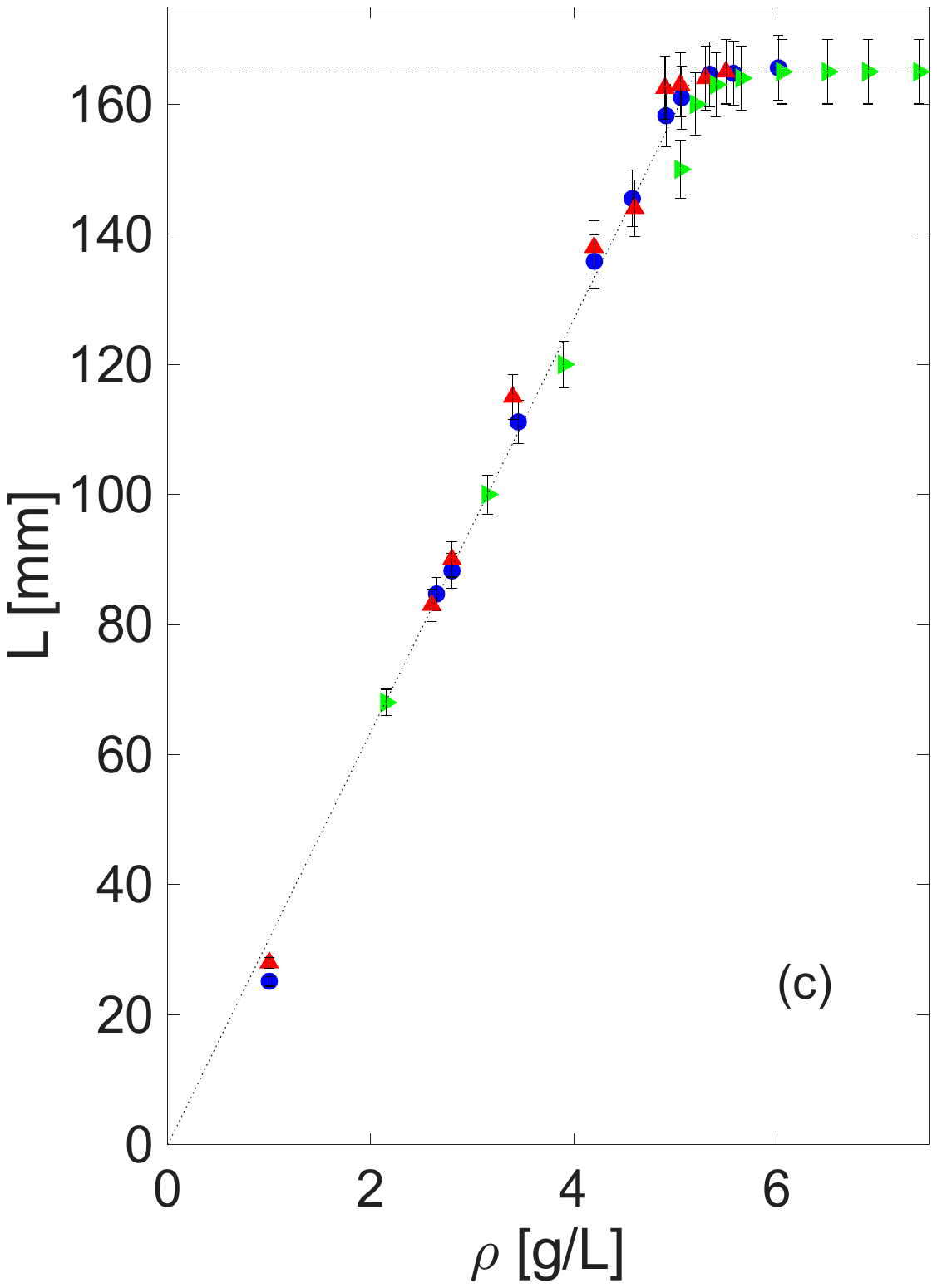}}
\caption{(a) Sketch and (b) front view of the apparatus with time lapse imaging. The sample comprised $12.7$ g of dry hydrogel beads in $2$ Litres of triply boiled tap water. The hollow sphere is $40\pm0.01$mm diameter and weighed $3.47$ g. The mid-plane of the sphere appears wider owing to lensing effects from the round container. The images are captured at $40$ second intervals. The equidistant separation of the time lapse images of the sphere highlights its constant speed. (c) The hydrogel sample surface as a function of hydrogel concentration shows a linear trend until the hydrogel packing reaches the water level (dashed line).}
\label{fig:intro}
\end{figure} 
 
\section{Overview}
\subsection{Constant speed motion}
It is perhaps surprising that a buoyant, rising sphere travels at a \emph{constant speed} towards a free surface through packings of non-Newtonian  hydrated hydrogels. Moreover,  previous observations highlighted time-dependency in the form of \emph{non-Newtonian creep}  in the motion towards a rigid end boundary in which the speed typically decreases with the square root of time~\citep{2022creepcontrol,2024creepconf}. The latter observations indicate the presence of significant time dependent, non-Newtonian behavior at these constant stress driving conditions.

Hence, observation of constant speed motion in the presence of a free surface is a novel and surprising result which reinforces the need for establishing a connection with the constitutive equations that govern soft sphere packings. Our current results suggests a concrete route towards necessary further theoretical work to describe the action of boundary conditions on the constitutive behaviour of hydrogel particle packings. Below, we offer a hypothesis that connects the drag force acting on the intruder to the exponential stress sensitivity of the rheology under different boundary conditions.\\

\subsection{Dependence on material properties and boundary conditions}
We use the observation of a constant velocity $U$ to characterize the motion of an intruder at a prescribed value of the stress and given boundary conditions. We show that the constant speed motion can be consistently \emph{changed} to a sub-linear time dependence of the displacement when a buoyant sphere rises towards a rigid lid resting on the upper boundary of the hydrogel suspension. We therefore systematically explore different boundary conditions of the hydrogel suspension to find that constant speed motion is found for all intruder motions towards a free boundary, including a  boundary formed with a heavy liquid below the hydrogel suspension, towards which the intruder is \emph{sinking}. 

Particle image velocimetry is used to show that the deformation field of the hydrogel suspension is highly localised around the intruder. Despite the localization of the flow, the free surface deformation observed when an intruder is far from it is substantial. We find the effective displacement rate of intruders inside the material to be exponentially dependent on the stress created by the intruder, which is consistent with our earlier work. This allows us to further probe the microscopic mechanisms behind the hydrogel packing rheology. The exponential stress dependence suggests Arrhenius or Eyring-type~\citep{eyring1935activated} mechanisms to be a leading factor in the dissipative mechanisms of the material, albeit that they occur at lower stress levels than the traditional pressure sensitivity of ordinary viscous fluids~\citep{bridgman1925viscosity}. However, the dependence on temperature of the hydrogel packing flow behavior is observed to be significantly greater than that of water alone. This finding indicates  the effective flow behavior of the suspension is not dominated by the solvent viscosity  e.g. via a local mechanism such as lubrication layers between the particles, as these would scale sinking or rising speeds in proportion with the water viscosity. 
While further rheological characterization of the bulk of the materials remains essential to understand the complete (microscopic) material behavior of hydrogel suspensions, we here focus primarily on the effects of the boundary conditions on intruder dynamics. As suggested in Sec.~\ref{sec:theory}, the boundary effects explored here only depend on the previously established exponential rate dependence of dynamics in hydrogel suspensions, yet are largely insensitive to the details of flow curve characterization, which we leave for future work. 

\section{Experimental set-up}
A front view of the apparatus used in the majority of the experiments is shown in Fig.\ref{fig:intro}.  It comprises a $130$~mm o.d., $200$~mm long  plexiglass cylinder which has a $3$mm thick wall. The cylinder is aligned vertically in the laboratory frame of reference and is mounted on  a $10$mm thick aluminium base which has a machined $129$mm diameter circular groove containing an 'O' ring. Silicone sealant is used to attach the cylinder to the base. The cylinder contains two litres of triply boiled Oxford tap water. Boiling is used to reduce the amount of dissolved gases in the water (see for example \citep{degassing}) and hence suppresses the formation of bubbles over the periods of the experiment, which were typically months. It is observed that  bubbles can significantly affect the reproducibility of the experimental results. This is a result of the effect of compressibility and surface tension on the packing mechanics~\citep{2022creepcontrol} and the relatively high Laplace pressure in small bubbles in comparison to the other stress scales that apply in the material. Repeatable results are obtained over the entire period of the experiment using the sample preparation protocol detailed below.

\subsection{Sample preparation} Sample preparation proceeds as follows. Measured amounts of dry hydrogel powder are slowly mixed with the water, the sample is gently stirred and left to settle for two hours between additions of dry particles. The particles swell as they absorb water and  sink to the bottom of the cylinder as they are slightly more dense than the fluid. The hydrogels absorb the water efficiently and form a layer of swollen particles at the bottom of the cylinder with a clear layer of fluid above. Measurements of the depth of the hydrogel layer plotted as a function of the added mass of dry powder is linear as in previous experiments~\citep{2022creepcontrol}. This finding is consistent with the hydrogel particles being essentially frictionless spheres~\citep{gong2006friction, workamp2019contact, Shakya2024}, which settle to a random close packing. The height of hydrogel particle interface reaches the surface of the water at a hydrogel concentration $C \approx 6$ g/L where $C$ is  the weight of the dry powder divided by the total volume of water in grams per Litre (g/L). A plot of the  packing height versus concentration for three separate samples is given in Fig.~\ref{fig:intro}.  The repeatability of this measure over a number of experiments is worthy of note since the creation of each sample takes several days. Once the layer has reached the surface, the addition of further quantities of dry hydrogels does not produce any detectable change in the layer height and presumably there is a small compression of the particles throughout the layer~\citep{2024creepconf}. The average size of the swollen particles is measured to be $\sim 1$ mm. 

The value of $C$ used in the majority of the experiments reported here is set to $6.35$ g/L. We choose to use the concentration $C$ to set the properties of the sample. As mentioned above, this is a robust control parameter which produces repeatable experimental results. The more common parameter of volume fraction used in studies of \emph{suspensions} of hard spheres \citep{pouliquen} is difficult to  estimate with soft materials. Recently, it has been shown that accurate estimates of the volume fraction of hydrogels in suspension \citep{swede} can be made reliably when specialized imaging equipment, such as X-ray facilities are available.

\subsection{Experimental Methods}
The motion of the sphere is tracked by capturing images using a computer-controlled camera at set intervals ranging from $5$ to $100$ seconds depending on the estimated speed of the ball. In the case shown in Fig.\ref{fig:intro}  the images are taken at $40$ second intervals using a Nikon $400$ digital camera. The recorded images are stored on a PC using the software DigiCam and are processed offline using the software ImageJ. The position of the sphere is estimated from the calibrated sequential still images and the speed is estimated from the resulting distance versus time plots.

The majority of the experiments were performed over a period of approximately one year in a laboratory where the measured temperature of the sample is $19.0 \pm  1.0$ °C. Over some periods the temperature varied outside of this range and the effects of this are reported below. However, the majority of the data given below is collected with the temperature within the given range. We will see below that this temperature window gives $\sim 10\%$ variation of measured travelling speeds, which is insignificant for the effects considered, given the large range of travelling speeds explored via the variation of buoyancy stress and intruder size.

Experiments were also performed at the University of Amsterdam, under standard laboratory temperature control. In these cases,  a suspension of hydrogel spheres was made inside a square plexiglass box of dimensions $25\times25\times40$~cm following our established protocol using dry hydrogel spheres with triply boiled Amsterdam tap water at a concentration of $6.09$ g/L unless otherwise noted. To initiate the experiments, a rod with a hoop was mainly used to place the sphere at the bottom, and in other cases a small magnet was placed inside the ball. An external magnet was used to hold and release the ping pong ball from the bottom of the container. No significant differences were found between the results obtained using these methods of control and release.

In the Amsterdam experiments the software TroublePix was used with a Lumenera Lt225C high speed camera to record a video of the motion at a frame rate of $71$ frames/second. Still images were captured using a Nikon D5200 camera with ball tracking  imaging performed using  MaxTraq and Matlab. The flow field was measured using Particle Image Velocimetry (PIV). In this, the plane of motion of the  sphere was illuminated by two vertical green laser sheets. Spatial averaging which co-moves with the intruder was used. Matlab integrated software ``PIVlab'' was employed to analyse the sequences of images.

\subsection{Upper Boundary Conditions}
The cylinder is sealed at the top by a tight-fitting lid containing a neoprene 'O' ring. The sealed container helps to minimise any evaporation of the water. There is a $\sim 1$ cm gap between the upper surface of the sample and the underside of the external lid. The upper boundary of the sample is a free surface for $\sim 20\%$ of the experiments and a printed plastic lid is placed on the surface of the sample for the remainder. The lid is  3D printed in plastic (PLA), hence it is partially aerated and weighs $\sim  50.3$ g. Its bottom surface is smooth since it is formed in contact with the heating element of the 3D printer, and is hence comparatively free from defects. It is placed on the surface of the hydrogel packing and a thin layer of water is squeezed  onto the upper surface. The results of an investigation into the effects of confinement \citep{2024creepconf} indicate a lid of this weight will produce a minimal contribution to the bulk stress on the material. 

Finally, a third upper boundary condition is imposed using a steel mesh placed on the surface of the sample. The stainless steel mesh was circular and contained an array of approximately square holes formed by $0.5$~mm wires which cross orthogonally. Meshes of this type are used in devices such as cafetieres. It has a diameter of $100$~mm, thickness $1.0$~mm and a pore size of $0.2$~mm and weighs $\approx 3.6$~g. Hence water may pass through the mesh easily, while hydrogels are held in position by the mesh. The surface of the mesh is  very rough, providing a no-slip boundary condition for the hydrogel particles.

A time lapse set of images are superposed on a frontview of the apparatus in Fig. \ref{fig:intro} to give an impression of the upward motion of a ping-pong ball. To initiate the experiment a buoyant ping pong ball is pushed down to the centre of the base of the cylinder using a $1$ mm diameter metal rod which has a hoop at one end. Smaller spheres were manipulated with a similarly shaped $0.5$ mm diameter rod. A settling period of $10$ minutes is allowed before the ball is released on the centre line and the hoop is carefully moved to the side of the cylinder. The timescale for the insertion and release of the sphere are much shorter than the transit time through the sample.

\subsection{Intruder specifications}
Ping pong balls are manufactured from Acrylonitrile Butadiene Styrene (ABS) plastic, have a diameter of $40\pm0.1 $mm and each weighs $2.73\pm 0.2$ g. Their size, sphericity and weight are strictly controlled by the International Table Tennis Federation~\citep{ITTFballstandards} and the dimensions of each ball are checked using a vernier gauge. The diameter ratio of the sphere to container $ d/D \approx 0.32 $ where $d$ is the diameter of the sphere and $D$ is that of the cylinder. Changing the buoyancy of a ping-pong ball is relatively straightforward, as water can be injected into the ball using a hypodermic needle, the injection hole sealed and the weight of the ball measured using an electronic balance. The buoyancy force can then be estimated using the difference in mass between the sphere and the displaced water. We assume that the stress $\sigma_S$ that an intruder exerts on the hydrogel suspension is its effective weight divided by its projected cross sectional area. Note that this is a somewhat different calculation of the stress exerted on a Newtonian fluid, which is integrated over the intruder's entire surface. However, the difference is a prefactor of order 1.

The main set of experiments are performed using ping-pong balls with a hydrogel sample with $C=6.35$ g/L.  A second set of experiments were carried out using a hollow $20$mm diameter polypropylene ball (Precision plastic ball co.). Injecting water and weighing is again used to change and estimate the buoyancy. The value of $C$ is $6.0$ g/L. A third set of experiments are performed with hollow  balls made from a range of plastics (polyamide, polypropylene, PTFE, polyethylene) and had diameters $25.0, 20.0, 19.0$mm. These hollow plastic balls were weighed using an electronic balance and the weight was used to compute the buoyancy stress. The value of $C$ used was $6.15$ g/L for these experiments. Finally, in one set of experiments, a glass sphere with a diameter of $25$mm was used.

\begin{figure}
\centering
{\includegraphics[clip, trim=1.5cm 7cm 2cm 6cm,scale=.6]{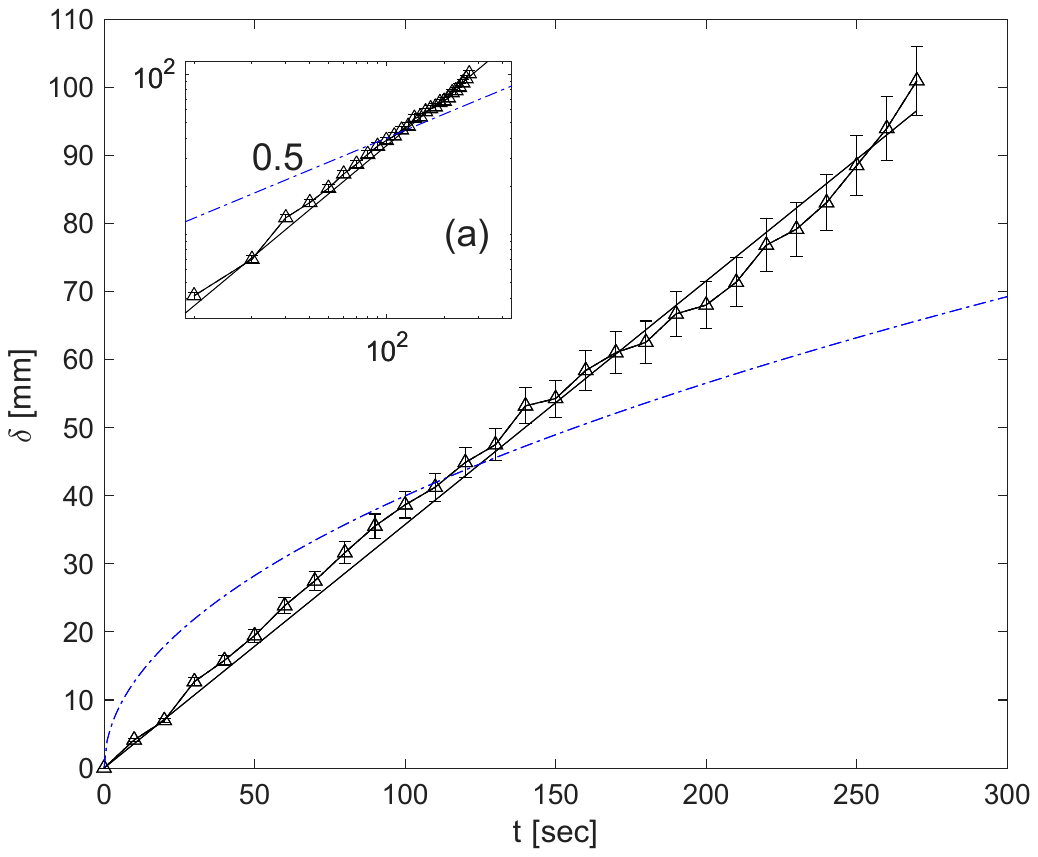}}
\caption{A distance $\delta$ versus time graph for  a sphere moving towards a free surface. The buoyancy stress is $\sigma_S \sim 210$~Pa. Error bars indicate a 5\% error due to imaging resolution limitations. The linear dependence of distance with time (solid line) is clear. For comparison with previous work we also indicate a square root time dependence (blue dash-dotted line). The inset (a) has the same axes but in double logarithmic scaling, uses the same symbols and highlights the square root $\delta\propto t^{1/2}$ motion and exponent of the blue dashed line.}
\label{resover}
\end{figure} 

\section{Motion of a rising intruder.}\label{section_rising_dynamics}                   

Distance versus time plots from a set of experiments with a free surface are shown in Fig.~\ref{resover}. It may be seen that the distance a sphere rises in this hydrogel sample is proportional to $t$ with a free upper surface on the sample. The inset Fig.~\ref{resover} confirms the linearity of the displacement dynamics on a double logarithmic scale. The constant speed motion is in stark contrast with previous square-root time dependence observations~\citep{2022creepcontrol, 2024creepconf}. We now present the results of a systematic investigation into this constant speed motion.

\begin{figure}
\centering
{\includegraphics[scale=0.5]{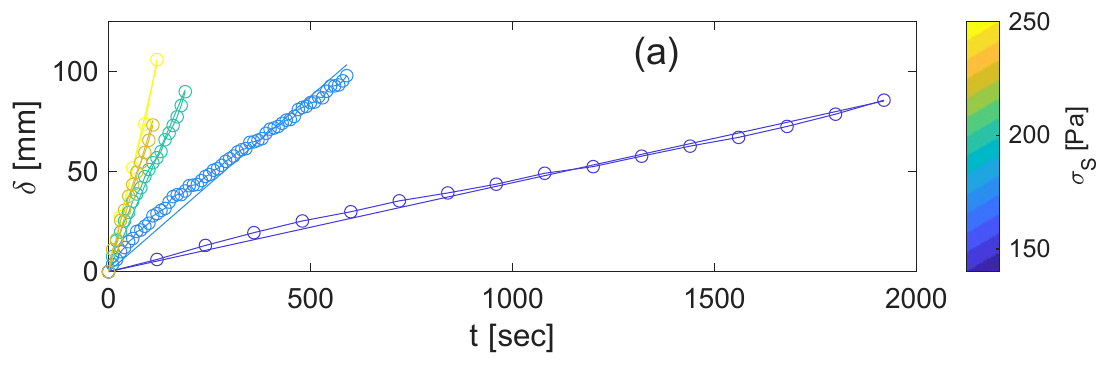}}
{\includegraphics[scale=0.5]{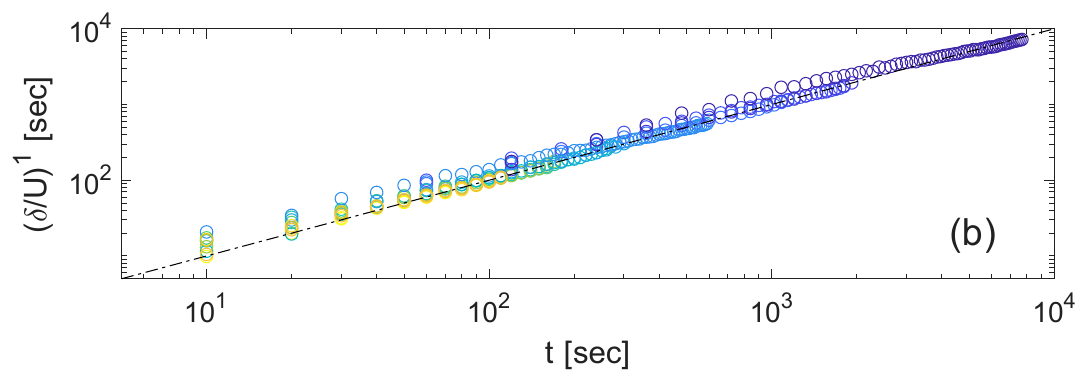}}
{\includegraphics[scale=0.5]{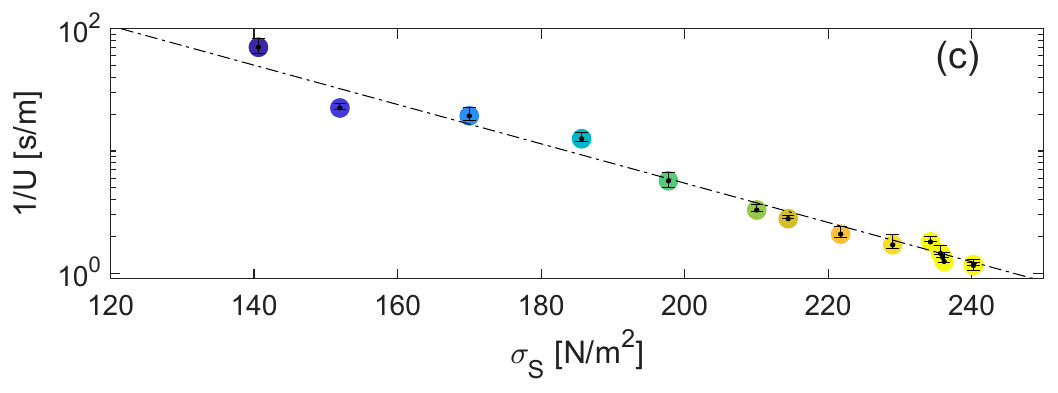}}
\caption{(a) Distance versus time plots for ping pong balls rising in a sample with a free surface. The rise speed depends on the weight of the partially filled sphere where the weight ranges from $0$ to $17.3$ g. This corresponds to a buoyancy stress $\sigma_S$ of $\approx 240$ to $\approx 130$ Pa. The lines indicate straight line fits to the data. (b) The displacement data can be rescaled by dividing out the terminal velocity $U$ to collapse all data on a linear master curve with slope 1 (dash-dotted line) over three orders of magnitude. (c) Graph of the slope of the fit to each measured time history plotted as a function of the buoyancy stress $\sigma_S$. Note the plot is presented on a log-lin scale. The dash-dotted line represents an exponential decay with a decay constant of $27$~Pa. The error indicates the fitting error.}
\label{fig:risinglinear}
\end{figure} 

\subsection{Time Histories with a Free Surface}
\label{subsec:timehistory}

As discussed above, the \emph{positive} buoyancy of the ping-pong ball can readily be changed by adding small measured quantities of water to its interior. The time histories obtained over a range of buoyancies are shown in Fig. \ref{fig:risinglinear} (a). It can be seen that a straight line provides a satisfactory least squares fit to each time history with typically a 3\% root mean squared error, and hence there is a linear dependence of distance on time. A wide range of buoyancies was explored, from a sphere without added water ($\sigma_S =240$Pa) and the total rise time is $\sim 2$ minutes to one with $\sigma_S = 130$ Pa, where the sphere takes $\sim 2$ hours to rise through the sample. We show all the $\delta(t)$ behaviour for the experiments with ping-pong balls in a sample with $C=6.35$ g/L and a free upper surface in Fig.~\ref{fig:risinglinear} (a). 

A fixed terminal velocity $U$ is found for each value of $\sigma_S$, which enables us to rescale all the displacement data acquired for one sample onto a master curve that follows $\delta \propto t$. We show the result of this analysis in Fig.~\ref{fig:risinglinear} (b). An excellent collapse over three orders of magnitude confirms the constant speed dynamics of an intruder rising towards a free surface. 

The collapse of the data provides an important insight. It is possible that the slowing down  associated with  $\sqrt{t}$ dependence observed previously~\citep{2022creepcontrol,2024creepconf} could arise from several sources. For example, a gradient in the properties of the prepared sample resulting from concentration variations, the effects of hydrostatic pressure, or other gradients because of  the sensitivity of the viscosity to packing density around the random close packing limit~\citep{shewan2015viscosity}. The constant speed motion, instead, highlights that carefully made concentrated samples of hydrated hydrogels do not display such spatiotemporal density variation effects induced by sample preparation. However, below we will show that the exponential compression sensitivity of the packing can be linked to the time dependent displacement dynamics.

A graph of the terminal velocities extracted from the time histories presented in Fig.~\ref{fig:risinglinear} (a), (b) versus buoyancy stress $\sigma_S$ is given in Fig.~\ref{fig:risinglinear}(c). The dependence is exponential as shown by the linear fit to the data, since the ordinate has a logarithmic scale. {We choose to show $1/U$ as this concept is most consistent with the prefactor $D$ used in earlier work~\citep{2022creepcontrol, 2024creepconf} and hence aids comparison. Hence, $1/U~\propto~\exp(-\sigma_S/\sigma_0)$ with $\sigma_0$ a constant that depends on the hydrogel density, among other factors. This exponential dependence of propagation speeds also depends on packing fraction \citep{2022creepcontrol} and global confinement stress~\citep{2024creepconf}, which is consistent with material response previously reported for these materials.

\section{End boundary conditions}\label{end_boundary_conditions}
All the data we have discussed thus far displays a linear dependence of distance on time for spheres moving towards a free boundary. As discussed in the introduction, previous results for the motion of heavy spheres through hydrated hydrogels ~\citep{2022creepcontrol,2024creepconf}, which correspond to cumulative 200+ measurements, show $\delta \propto t^{1/2}$. The main difference between the two configurations is that in the present study, the ping-pong balls rise towards a free surface, whereas the heavy spheres and cylinders used in earlier work descended towards a rigid boundary. We thus identify several additional boundary conditions that need to be explored to investigate the role of end boundary conditions:

\begin{itemize}
    \item[I] An intruder rising towards a rigid boundary.
    \item[II] An intruder sinking towards a rigid boundary.
    \item[III] An intruder sinking towards a free boundary.
   \end{itemize}


\begin{figure}
\centering
{\includegraphics[scale=0.5]{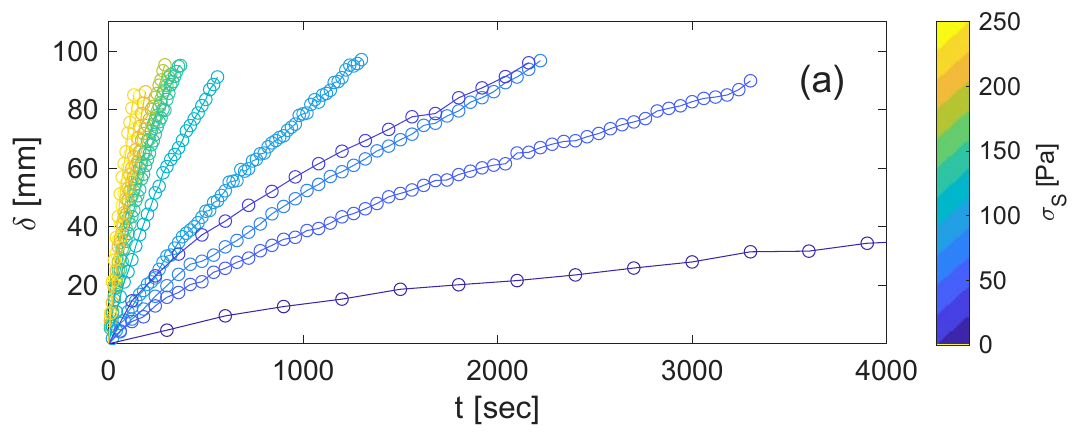}}
{\includegraphics[scale=0.5]{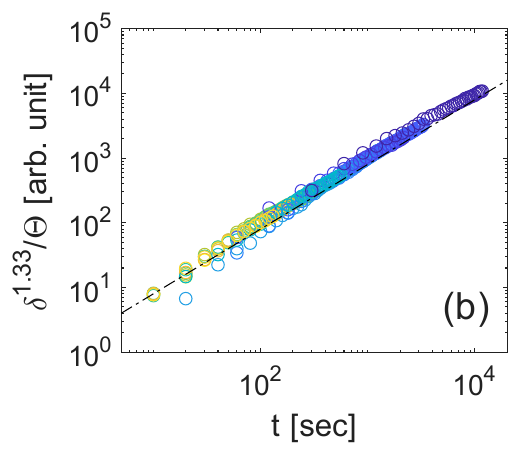}}
{\includegraphics[scale=0.5]{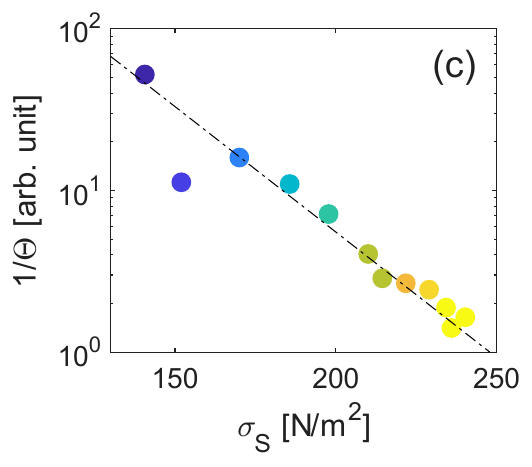}}
\caption{(a) Time histories of  ping pong balls over a range of buoyancies rising through a hydrogel sample towards a rigid lid. The  sample used was the same as used to obtain data for Fig.~\ref{fig:risinglinear}. The nonlinear data is fitted with the  function $\delta(t) = \Theta't^\alpha + c$ with $\alpha= 0.73$. (b) rescaling $\delta(t)^{1/\alpha}/\Theta$. The dash-dotted line has slope 1. (c) $1/\Theta$ depends exponentially on $\sigma_S$, as indicated by the dash-dotted line. The line represents exponential stress dependence with a stress scale of 28 Pa, similar to the function as used in Fig.~\ref{fig:risinglinear}b.}
\label{fig:risingnonlinear}
\end{figure}

\subsection{\rm{[I]} Sphere moving towards a rigid lid.}\label{towards_lid}
We perform a sequence of experiments in which a printed plastic lid sits on top of the swollen hydrogels, while an intruder moves towards the lid.  The sample used as before has $C=6.35$ g/L and the buoyancy of the sphere is adjusted by adding or subtracting water to and from the interior of the ping-pong ball as described previously.
At the same value of $C$ and buoyancy, the sphere rises with a linear distance-time relationship with a free surface. However, it is sublinear, when the lid is present. Note that a dependence of distance travelled on $\sqrt{t}$ is qualitatively different to the end effects in viscous Newtonian flows \citep{tanner}.  

The observed sublinear dependence suggests a systematic exploration of the effects  of a lid on $\delta(t)$ behavior is required. The results of experiments with ping pong balls of various buoyancies are shown in Fig.~\ref{fig:risingnonlinear} (a). We observe a clear sublinear time-dependent behavior for the motion of the sphere. In order to extract the time dependence of the displacement of the rising sphere, we again fit to each data set 
\begin{equation}
    \delta(t) = \Theta'(\sigma_S)t^\alpha + c \label{eq:nonlin}
\end{equation}
Here, $\Theta' = \Theta^\alpha$ is a prefactor set by the speed of motion. The prefactor is potentially dependent on $\sigma_S$. $c$ is a small offset term that represents displacement uncertainty. We fit the prefactor $\Theta'$ and $\alpha$ independently and extract $\Theta$ afterwards to avoid a correlation between the two fitting parameters. 
We  verify which rescaling coefficient provides the best collapse of the data; plotting $\delta^{1/\alpha}/\Theta$  shows a linear time dependence. We find that $\alpha = 0.75$ provides the best collapse of the data and show this result in Fig.~\ref{fig:risingnonlinear}b.  We verify the exponential dependence of prefactor $\Theta$ that we use to characterise the strength of the sublinear behavior in Fig.~\ref{fig:risingnonlinear}c. Since the same hydrogel sample was used, we can compare the exponential dependence on the buoyancy stress for the sublinear case with the case of the sphere rising towards a free surface as discussed in Sec.~\ref{subsec:timehistory} (dash-dotted line). The data for the sublinear case clearly follows the same scaling of $\Theta \propto \exp(-\sigma_S/\sigma_0)$, with the same $\sigma_0 = 23 \simeq 28$ Pa as used for the free surface case. We may not expect a perfect match in the prefactor, as $\Theta$ absorbs some of the physical prefactors that we could separate out in the constant velocity case.  We conclude that the motion of a sphere towards a rigid boundary is sublinear by way of contrast with the motion towards a free surface, which is linear.

\begin{figure}
\centering
{\includegraphics[clip=true, trim=1.5cm 6cm 1.5cm 5cm,scale=0.5]{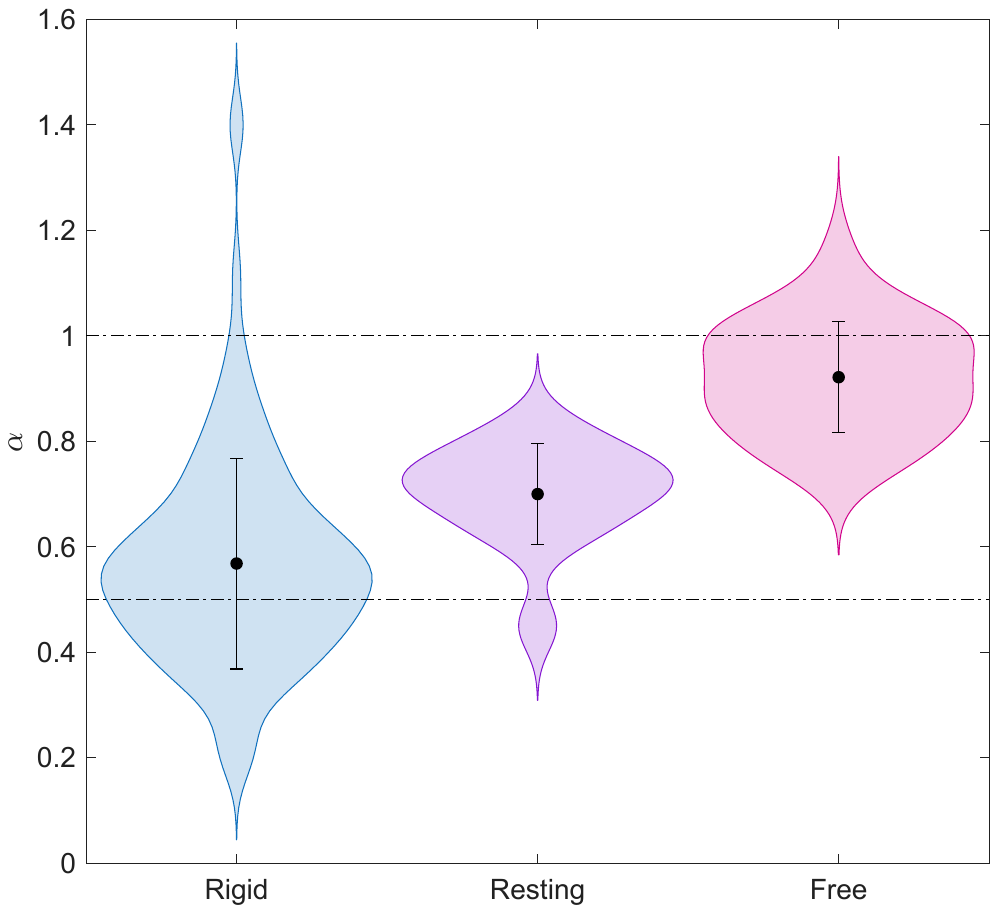}}
\caption{Violin plots of all exponents $\alpha$ extracted from a fit of the displacement data. The data labelled ''Rigid'' corresponds to  data from previous studies. The data labeled ``Floating'' corresponds to observations of rising spheres that move towards a floating  rigid lid. N.B. Superposition of images captured as the spheres approach and touch the lid indicate that there is no detectable movement of the lid. The data labelled ``Free'' shows the data of an intruder rising towards a free  surface. For each, the average exponent observed is indicated with a dot, the standard deviation of all measured exponents are shown as error bars, and the distribution of exponents observed are used to construct the violin plot. Dash-dotted lines highlight two standard exponents 0.5 and 1 for comparison.}
\label{fig:exponentstats}
\end{figure}


\subsection{\rm{[II]} Sphere moving towards a rigid base --- comparison with previous work.}

In previous work~\citep{2022creepcontrol,2024creepconf} we performed experiments with sinking intruders that moved towards the rigid base of the container. These additional data enable us to compare the observed time dependence with the cases reported here. We extract the data from the open repositories, e.g.~\citep{data_SM} and perform the same fitting procedure to estimate the $\delta(t)$ behavior for these data sets. To allow for a systematic comparison of our results, we publish our kinetic data and analysis scripts in an open repository~\citep{DijksmanEtAl_2025}.\\ 
We can extract the exponent for which a power law model $\delta(t) = \Theta t^{\alpha} + c$ provides the best fit. We note that there is no canonical definition of a ``best'' fit, as fit quality as measured by a reduced $\chi^2$ depends also on weights associated with data, and even with an optimal $\chi^2$, systematic deviations in the residual of the fit are neglected. Our choice for ``best'' is hence simply the best goodness of fit estimate as returned by the MATLAB-based fitting routine.

Applying this  standard across all data sets from two previous publications and the current one, yields the results as shown in Fig.~\ref{fig:exponentstats}, with over 300 $\delta(t)$ sets fitted. The statistics of the distribution of $\alpha$ for each case of ``rigid'' boundary, ``resting lid'' and ``free'' surface is indicated with a violin plot, including a separate scatter plot for the mean and standard deviation of the distribution (error bar). We observe a clear clustering of exponents around 0.5 for all the previously published data, in which an intruder was sinking towards a rigid base. In the present study, some data was obtained with an intruder rising towards a rigid lid sitting on the hydrogel packing; the exponents for these experiments center around 0.7. The exponents for the creeping flow measurements for intruders that move towards a free boundary clearly center around $\alpha=1$, confirming again the strong influence of the boundary on the flow behavior of the hydrogel suspension. Lastly, for the data sets of the previous investigations we  obtain a satisfactory collapse as obtained for the current data as shown in Fig.~\ref{fig:risingnonlinear} (b). The temporal resolution in our previous work was lower than used here, but the exponent which provides a satisfactory collapse for these datasets is $0.5$ for data from ~\citep{2022creepcontrol,2024creepconf} albeit with increased scatter. 

\begin{figure}
\centering
{\includegraphics[scale=0.5]{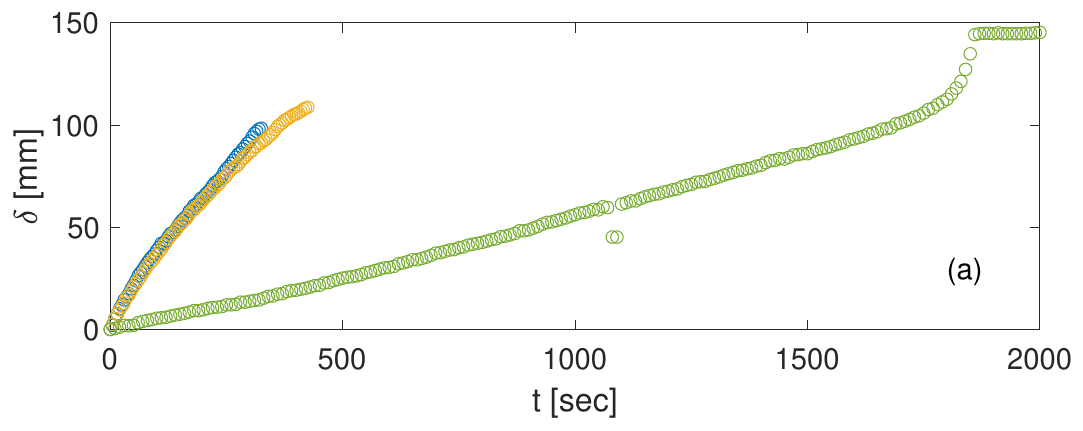}}
{\includegraphics[scale=0.5]{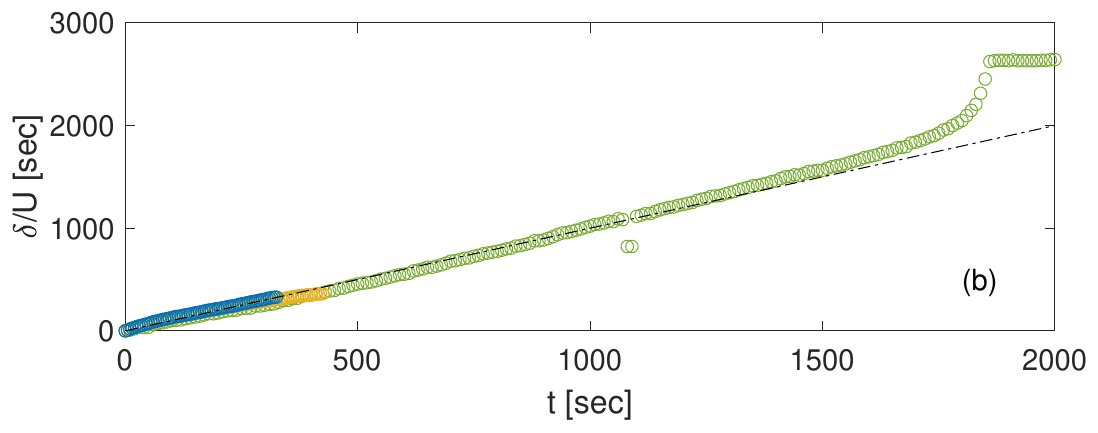}}
\caption{(a) Displacement of sinking spheres towards a fluid boundary. The top panel shows the data for three different sinking experiments: $C = 3.14$ g/L, a standard ping pong ball filled with water as intruder, with a diameter of 40 mm and mass of 43 g for the blue and yellow (repeat) datasets; $C = 3.21$ g/L, a glass marble as intruder, with diameter of 25 mm and mass of 20 grams for the green dataset. (b) Rescaling the data with their displacement velocity $U$ extracted from the linear parts of the data collapses all data on a master curve with slope 1 (dash-dotted line). The outlier data points in the green dataset are due to an illumination-induced tracking error.}
\label{fig:sinkinglinear}
\end{figure}

\subsection{\rm{[III]} Sinking towards a free, fluid boundary.}
To investigate whether the constant speed behavior exists for heavy spheres that \emph{sink} towards a ``free'' boundary, we perform experiments on sinking spheres moving towards a boundary formed under the hydrogel suspension by placing a heavy immiscible fluid at the bottom of the container. In these experiments, the hydrogel suspension $(C= 3.14$ or $3.21~gL^{-1})$ was created on top of a $2$~cm liquid layer of Galden 135 fluid, with a density of $1.72~gL^{-1}$. The hydrogel suspension was made with deionised water, in which the hydrogel particles swell more easily, as additional ions in the water screen the electrostatic forces that induce swelling~\citep{hooper1990swelling, baker1995swelling, okay2000swelling}. This accounts for the lower value of $C$ used here in comparison with the typical values used in the ping-pong ball experiments. We confirmed that this lower value of $C$ produced a sample in which the entire water volume is densely packed with hydrogel spheres. 

First, a water-filled ping pong ball is used in the initial experiments. Hence, it is negatively buoyant in the hydrogel suspension, but positively buoyant in the heavy Galden liquid. The motion of the filled ping-pong ball achieves a constant velocity while fully immersed in the hydrogel suspension but slows to a standstill at the interface of the hydrogel suspension and the Galden. The observations are found to be repeatable. The second set of experiments involves  a glass sphere ($\rho = 2500$kgm$^{-3}$), which is negatively buoyant in both the hydrogel suspension $(C = 3.21 gL^{-1})$ and Galden. The hydrogel suspension used in the second experiment is slightly denser to set the sinking time to be consistent with previous observations. Despite the differences between the two sinking experiments, the sphere reaches a constant velocity as indicated in Fig.~\ref{fig:sinkinglinear} (a). Rescaling the displacement data with the terminal velocity of the respective suspension-intruder pairs collapses the data in a straight line with slope $1$, confirming  that even if the motion is in the direction of gravity,  an intruder driven by a constant stress moving towards a free boundary of a hydrogel suspension maintains a constant speed.

\begin{figure}
\centering
{\includegraphics[clip, trim=1.5cm 6cm 2cm 5cm,scale=.35]{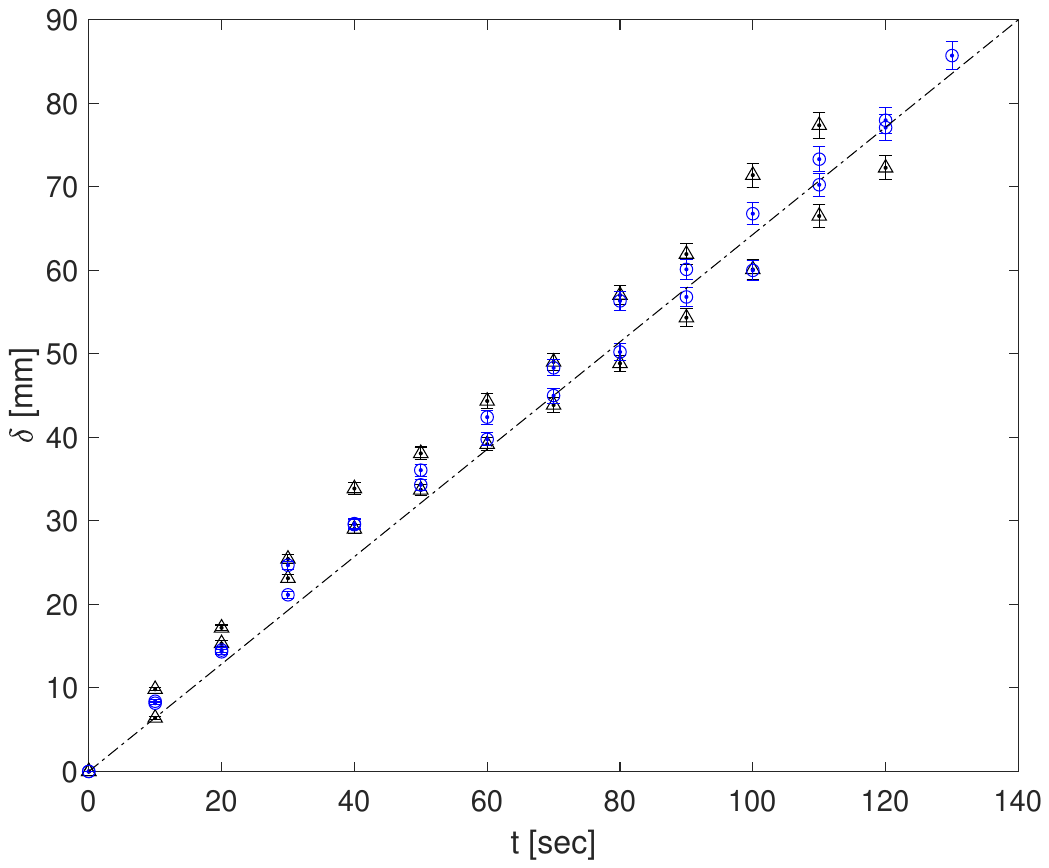}}
\caption{Four time histories with  $C=6.35$ g/L. Blue circles represent experiments with a free surface and black triangles represent experiments in which a metal mesh rests on the upper surface of the hydrogel packing. The straight dash-dotted line is a guide to the eye.}
\label{fig:mesh}
\end{figure}

\subsection {Rising towards a surface with a mesh}
Having considered a systematic variation of completely unimpeded and rigid boundaries and positive and negative buoyancies, we show four time histories in Fig.~\ref{fig:mesh} four  with hydrogel density $C =6.35$ g/L. Two of the time histories were measured with a free surface and two with a metal mesh which sits $\sim 1$mm below the surface of the packing. The mesh prevents the hydrogel particles at the surface from moving during the transit of the ping-pong ball, yet it is sufficiently porous that water flows relatively unimpeded. In contrast to the floating lid, the boundary is rough and hence suppresses radial motion of the hydrogels at the top of the packing; as we shall see below in Sec.~\ref{subsec:radial_displacement_field}, such radial motion is substantial in the case of the free end boundary. However, the mesh is also very light. The results indicate that the mesh has very little effect on the constant speed motion of the ping-pong ball. Moreover, the surface of the sample remains flat during the motion of the ball, indicating that the constant speed motion is not an end effect resulting from the imposition of  a  structure on the surface, suggesting that flexible rough lids will also not create sublinear motion. We cannot exclude the possibility that a smooth yet flexible boundary such as the surface tension boundary conditions from~\citep{2022creepcontrol} can nevertheless be essential to create sublinear motion.

\begin{figure}
\centering
{\includegraphics[scale=0.5]{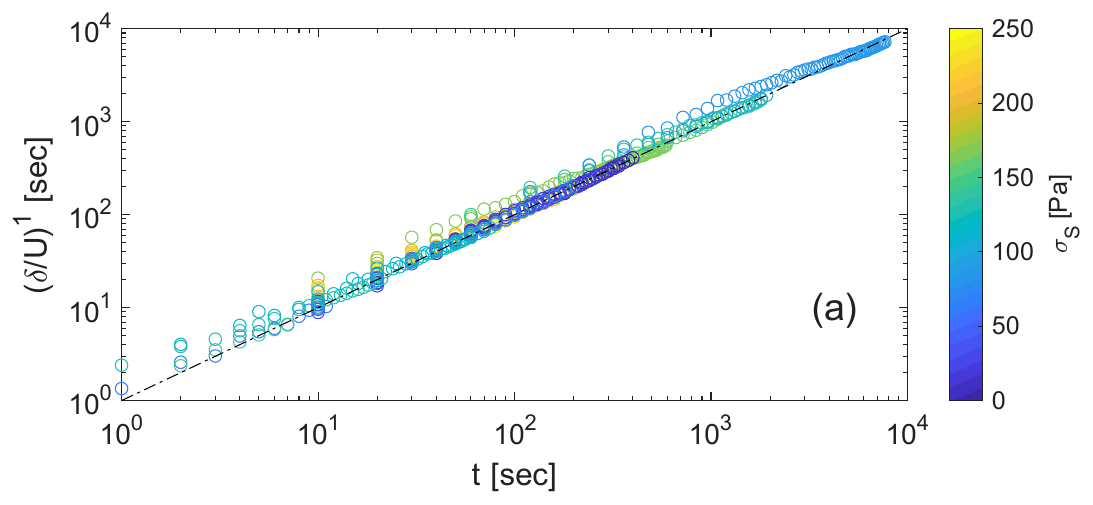}}
{\includegraphics[scale=0.5]{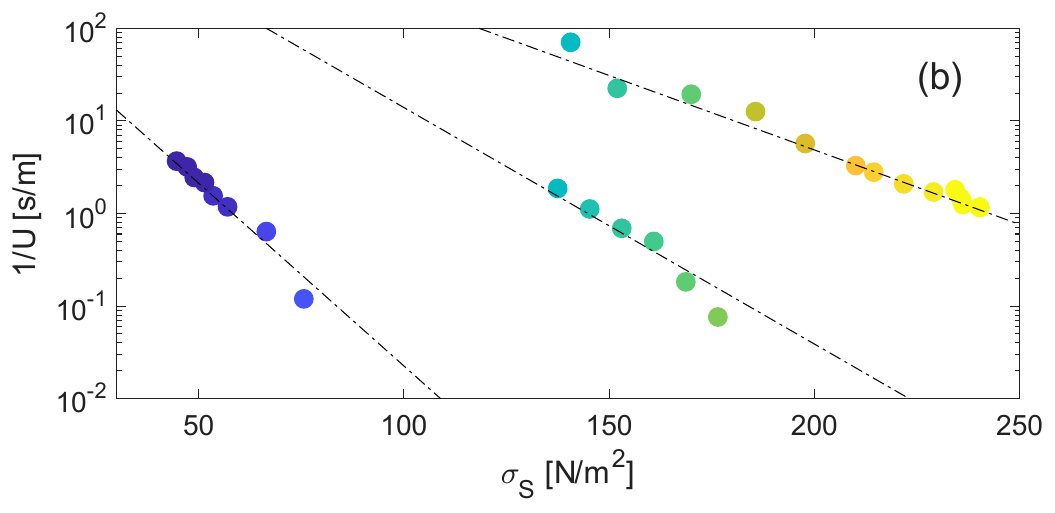}}
\caption{(a) Time histories obtained with ping-pong balls scaled using the fitted velocity $U$ for samples of hydrogels with $C = 6.35, 6.15$ and $6.0$ g/L with a free surfaces. (b) Log-linear plot of the effective viscosity based on $U$ of the hydrogel sample versus buoyancy stress $\sigma_S$ for cases with lid-free surface. The dash-dotted line indicates an exponential dependence on $\sigma_S$ with stress factor $\sigma_0$, which depends on the sample. We used $\sigma_0 = 27,17,11$~Pa for the different samples respectively. 
}
\label{fig:stressdep}
\end{figure}

\subsection{Effects of changes in $C$}
We now consider the consistency in our observations  to changes in particle concentration $C$ . Constant speed motion is  consistently observed for three different values of $C = 6.35, 6.15$ and $6.0$ g/L as shown in the results presented in Fig.~\ref{fig:stressdep} (a). For these  samples, there are good least squares fits of a straight lines with comparable accuracy to those reported above.  This indicates  a constant velocity approximation is an appropriate descriptor for the dynamics of the sphere rising towards a free surface. The terminal speed of the sphere can again be estimated from the slope of the least squares fitted line through $\delta(t)$ and an estimate of Fig.~\ref{fig:stressdep} (a) is made in each case. The fitted value of $U$   can be used to rescale the time history plots and a convincing collapse of the data is found as shown over four orders of magnitude. Each measurement indicates an exponential dependence of $U$ on $\sigma_S$. This trend is shown in Fig.~\ref{fig:stressdep}(b) plotted as a function of buoyancy stress $\sigma_S$. As expected, based on previous work~\citep{2024creepconf}, for each sample, the characteristic stress $\sigma_0$ of the scaling $1/U\propto \exp(-\sigma_S/\sigma_0)$ is slightly different.

We conclude that despite the constant speed motion of the sphere, the fluid medium is not Newtonian: there is an exponential drop in the resistance to motion as the buoyancy is increased. Specifically, the log-linear plot of the slopes of the $\delta$ versus $t$ plots shown in Fig \ref{fig:stressdep} indicates that the propagation speed of the sphere depends exponentially on the buoyancy stress with stress factor $\sigma_0$, which varies between $10$ to $23$ Pa. This finding is consistent with previous results for stress dependence of creep at both local \citep{2022creepcontrol} and global \citep{2024creepconf} levels for both spheres and cylinders. 

\begin{figure}
\centering
{\includegraphics[scale=0.5]{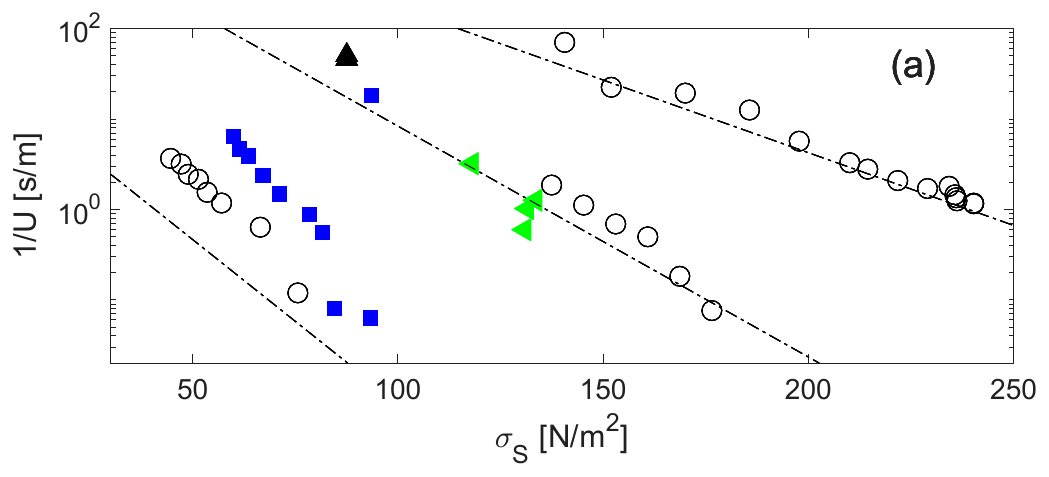}}\\
{\includegraphics[scale=0.5]{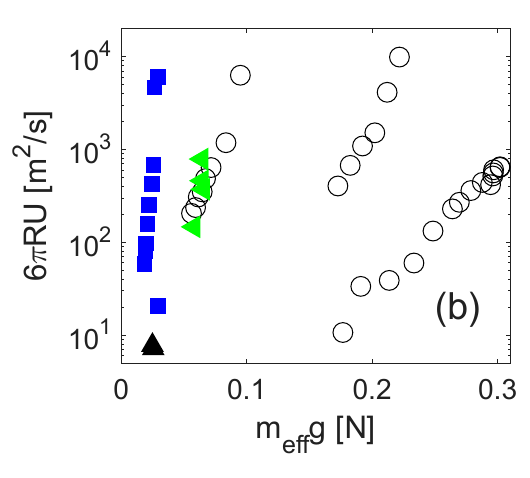}}
{\includegraphics[scale=0.5]{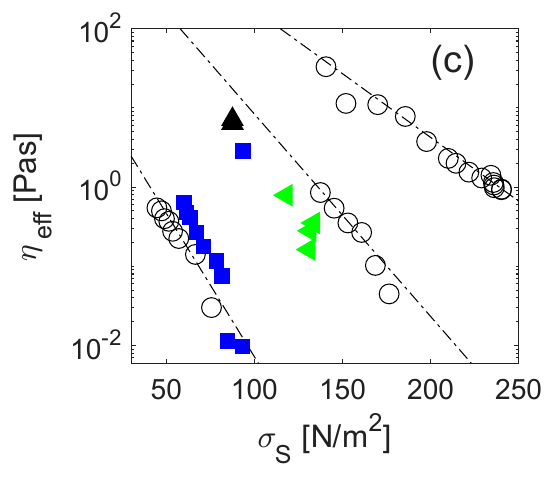}}
\caption{(a) Terminal velocity for different intruders and compared to ping pong ball data. Black $\triangle$: 19~mm sphere; green $\triangleleft$: 25~mm sphere; blue $\square$ 20~mm sphere. Open circles and the dash-dotted lines indicate the exponential dependence on $\sigma_S$ as shown in Fig.~\ref{fig:stressdep}. (b) Log-linear plot of a Stokesian drag force versus gravitational driving term, which should have provided a linear relation for a Newtonian sample (c) Log-linear plot of the effective stress-dependent viscosity calculated with Eq.~\ref{eq:effvisc} for all data from panel (a).}
\label{fig:stressdepcomp}
\end{figure}

\begin{figure}
\centering
{\includegraphics[clip, trim=1cm 8.5cm 1cm 7cm,scale=0.5]{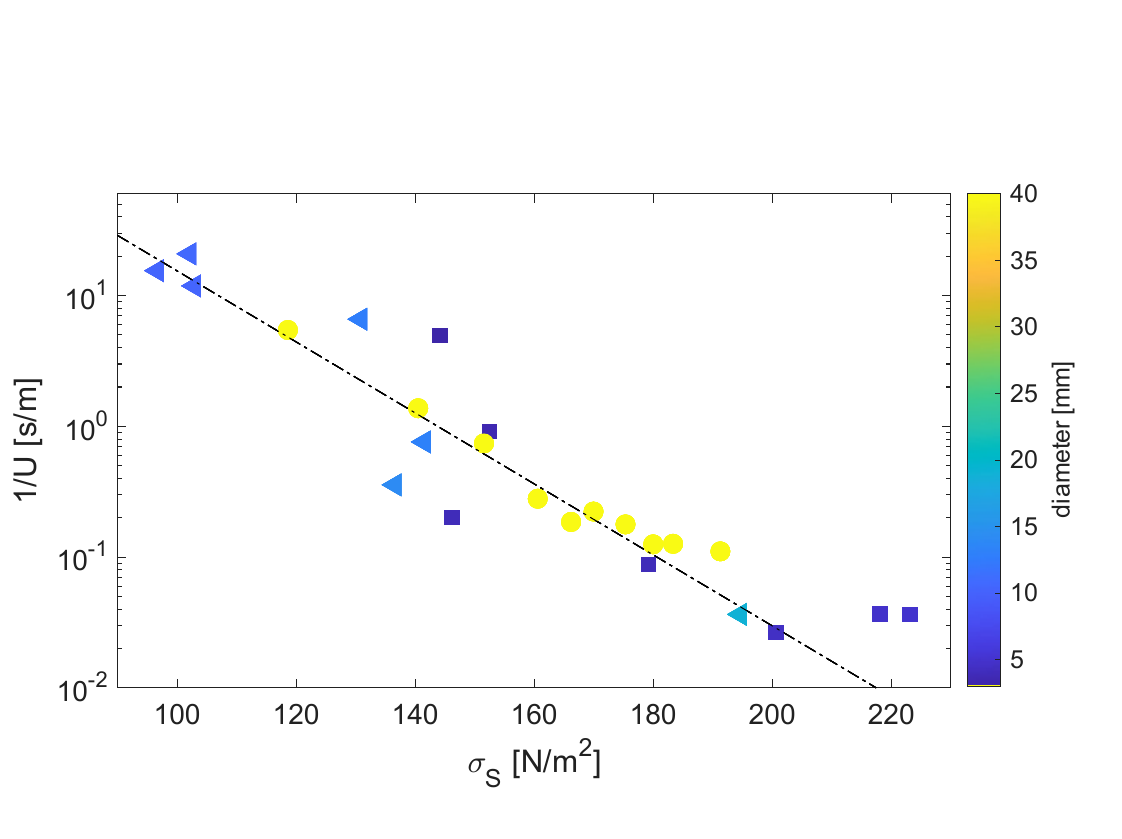}}
\caption{Sinking tests for very small spheres: glass spheres $\triangleleft$, stainless steel $\square$; ping pong ball $\circ$. The dash-dotted line represents an exponential stress dependence with decay constant $\sigma_0=16$ Pa. The color bar indicates the diameter of the spheres.}
\label{fig:smallspheres}
\end{figure}

\subsection{Results with smaller diameter spheres.}
Smaller diameter buoyant spheres are used to test the robustness of the results. These include a hollow plastic sphere of $20$ mm in diameter, which is filled with different amounts of water to reach the desired weight. Hollow PTFE, delrin and polypropylene spheres of $19$~mm, $20$~mm and $25$~mm are also employed. These intruders are used with the hydrogel sample at $C=6.15$ g/L. The hollow plastic sphere was also tested on the hydrogel sample with $6.0$ g/L. We find that the motion for the smaller diameters also have a terminal velocity that depends exponentially on the stress they exert on the hydrogel packing, after rescaling for the exerted stress on the cross section of the sphere, as shown in Fig.~\ref{fig:stressdepcomp}a. Results are in agreement with the data for the ping-pong balls over the entire range of buoyant stresses investigated and cluster in the same way for different hydrogel samples. We conclude from these results that finite size effects of the spheres are not significant, and that a intruder-stress driving mechanism must be at play.

In principle, the existence of a terminal velocity of the intruder enables the assignation of 
an effective viscosity to suspensions of particles in fluids~\citep{nNexp, Cheng}. Using this approach for \emph{suspensions}, one can extract an effective viscosity from the constant intruder velocity data by converting the directly measured terminal velocity $U$ of the rising sphere into a value with the units of Pas, using the experimental variables $R$ as the radius of the sphere, $\Delta\rho = \rho_I-\rho_h$ as the difference in density between the intruder and the average of the hydrogel medium respectively and $g$ for gravity. In the observed force balanced motion, one can then verify if $4\Delta\rho R^3g/3 = 6\pi\eta_{\rm eff}RU$, leading to:
 \begin{equation}
    \eta_{\rm eff} =\frac{2\Delta\rho R^2g}{9U}, \label{eq:effvisc}
\end{equation}
Stokesian dynamics would provide a linear relation $4g\pi\Delta\rho R^3/3 \propto 6\pi RU$, only affected by the constant effective viscosity of the sample. Fig.~\ref{fig:stressdepcomp}(b) shows that this proportionality is neither linear nor independent of the intruder size, so the samples are clearly non-Newtonian. However, in this interpretation, $\eta_{\rm eff}$ can still be the effective dynamic viscosity of the fluid, which may depend on the applied stresses $\sigma_L, \sigma_S$. We use Eq.~\ref{eq:effvisc} to convert the terminal velocities from all previous rising sphere experiments and show the results in Fig.~\ref{fig:stressdepcomp}(c). We observe an improved collapse of the data with respect to Fig.~\ref{fig:stressdepcomp}(a), providing a rationale to consider the terminal velocity as a proxy for the effective intruder-stress dependent viscosity of the fluid. Note that the constant velocity motion experiment yield effective viscosities that ensure that the effective Reynolds number $Re = \rho U R/\eta_{\rm eff} \ll 1$.\\

In order to explore possible finite size effects even further, we carried out experiments with small solid glass and stainless steel spheres with diameters ranging from 3 to 19 millimeters and density of $2,5$ and $7,8 kg m^{-3}$ respectively. We measure the terminal sinking velocity $U$ as a function of the intruder stress $\sigma_S$ both extracted in the same way as described above. We perform these tests in a new hydrogel sample with $C=6.0 g L^{-1}$. We also perform tests using  a ping pong ball to check the consistency of the new hydrogel sample. We find that the exponential stress dependence of the sinking velocity is in accord with those found for all the significantly larger ping pong balls. This is the case  even for the steel spheres which are only approximately a factor of three larger than the swollen hydrogels. The results are shown in Fig.~\ref{fig:smallspheres}. We note that sinking dynamics towards the rigid base is always at constant velocity for these small spheres, consistent with the arguments provided in Sec.~\ref{sec:theory}. As in the limit of intruder spheres that approximate the particles size, our results remain robust, we confirm that finite size effects are limited.

\section{Flow structure within the hydrogel packing}

\begin{figure}
\centering
\begin{minipage}[c]{0.6142\linewidth}
    \centering
    \includegraphics[width=0.4301\linewidth]{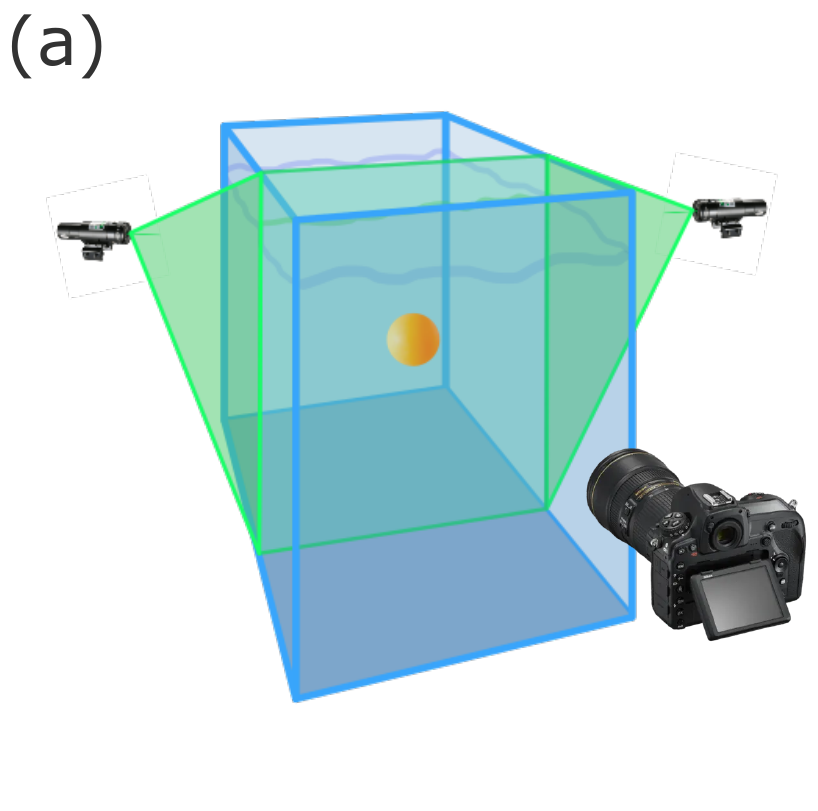}%
    \hspace{0.0062\linewidth}%
    \includegraphics[width=0.5121\linewidth]{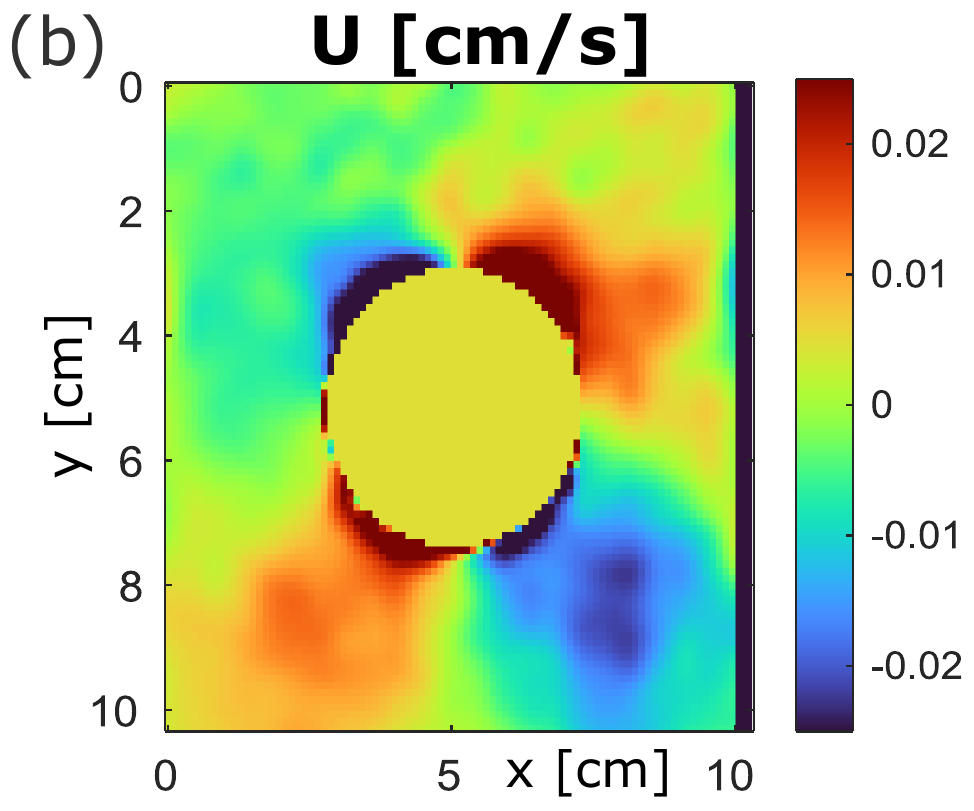}\\[1.5mm]
    \includegraphics[width=0.4832\linewidth]{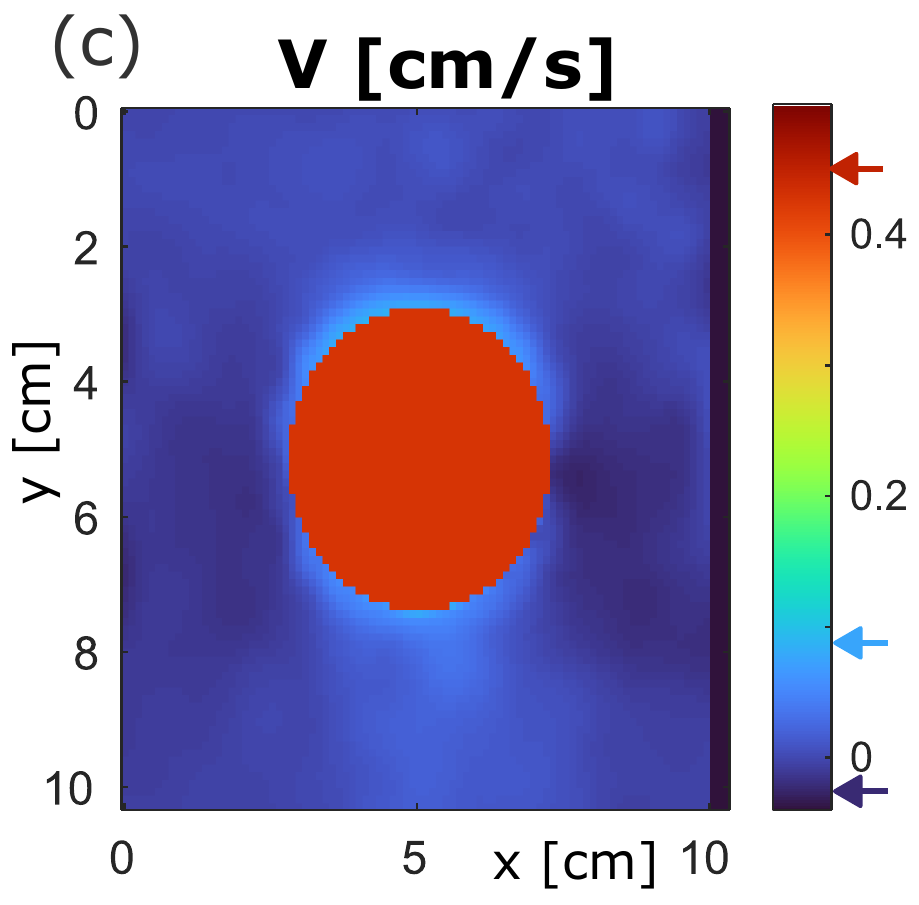}%
    \hspace{0.0062\linewidth}%
    \includegraphics[width=0.5106\linewidth]{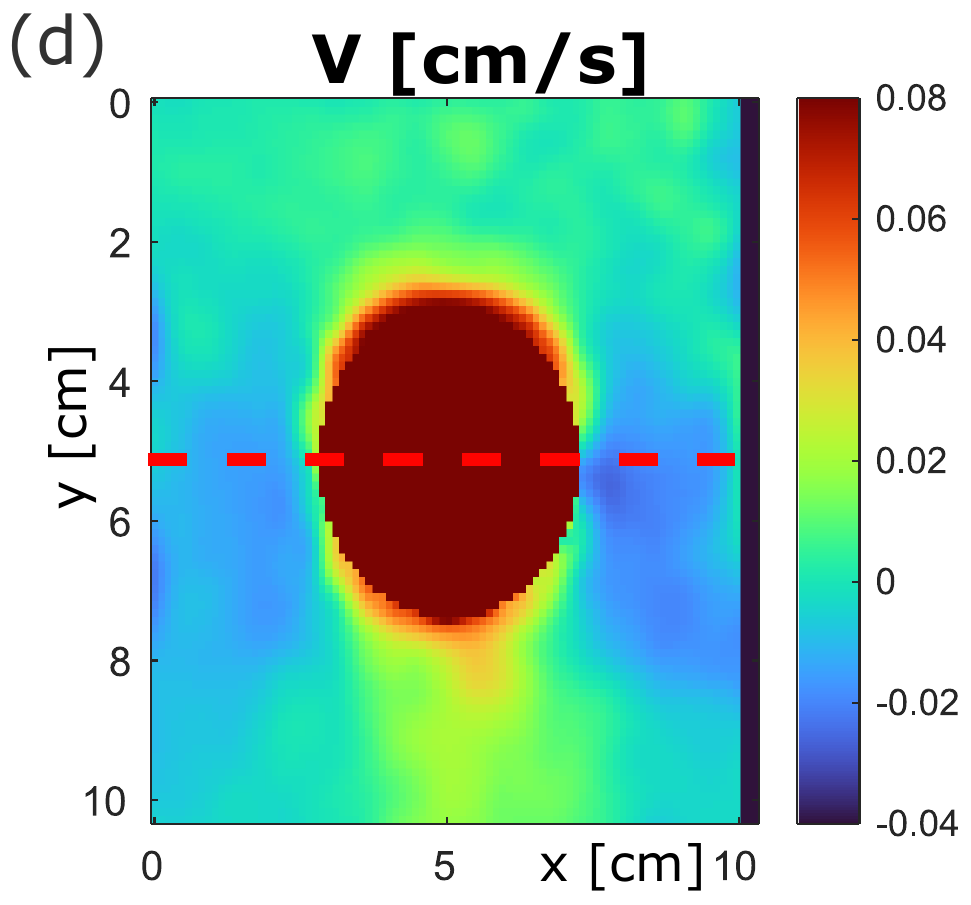}
  \end{minipage}%
  \hspace{0.0038\linewidth}%
  \begin{minipage}[c]{0.3820\linewidth}
    \centering
    \includegraphics[width=\linewidth]{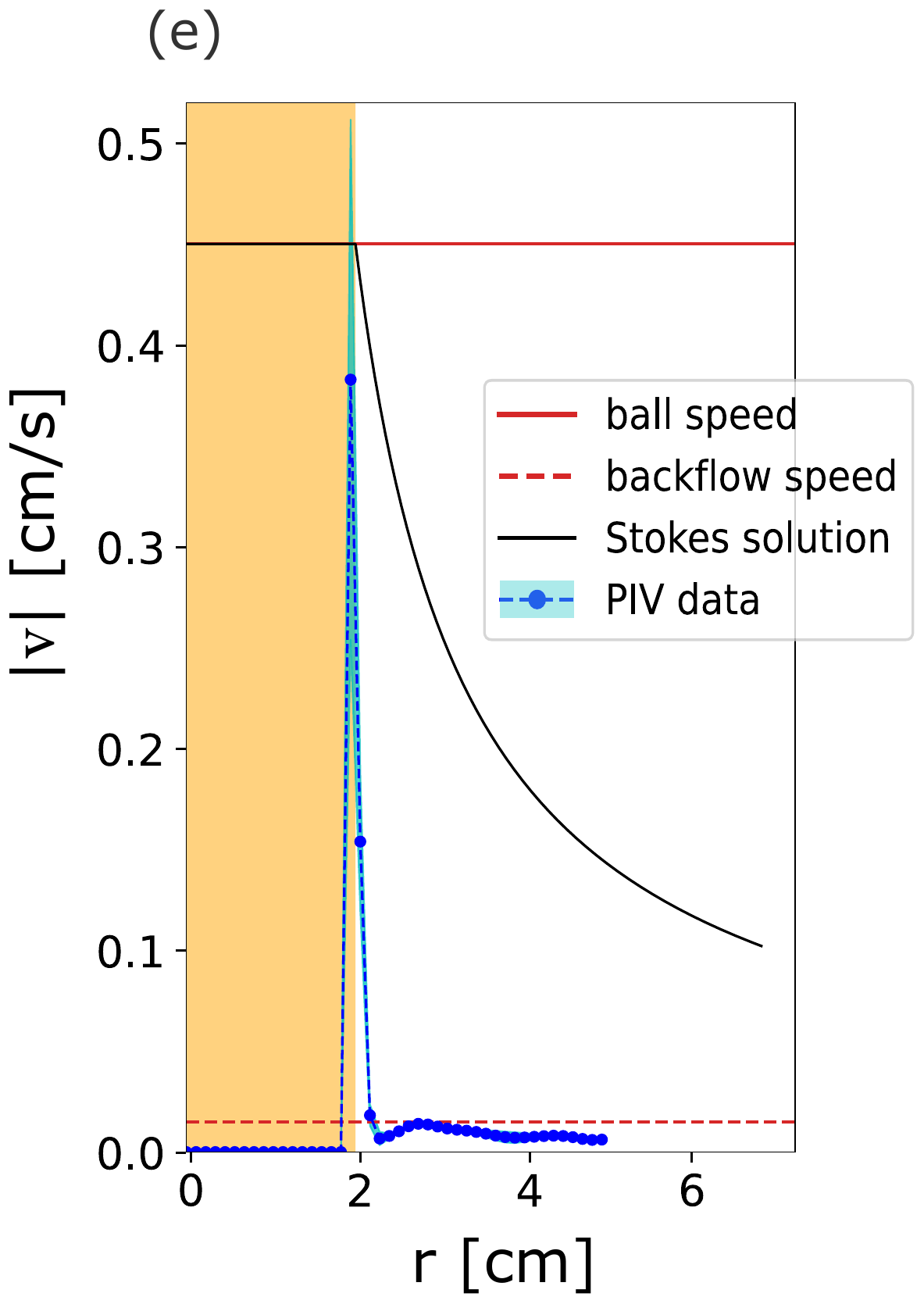}
  \end{minipage}
\caption{Results of PIV on the rising ball. (a) Sketch of the experimental setup. (b-d) Visualisation of the flow field: (b) the horizontal velocity component $U$ clearly shows a quadrupolar nature, with a slight vertical asymmetry; (c) the vertical velocity  of the ball, marked by the red arrow in the color bar on the side, is an order of magnitude bigger than the backflow component in the hydrogel motion $V$, marked by the dark blue arrow, and  a factor 5 bigger than the small clump of hydrogel particles dragged along by the ping pong ball; (d) details of the vertical variation of the flow revealed by constraining the color scale as compared to (c). The dashed horizontal red line represents the plane analyzed in panel (e), where we can observe a comparison of the PIV measure of $|v| = \sqrt{V^2 + U^2}$ at the midplane of the sphere with respect to the expected radial Stokes solution at that point. The sensitivity of the measurement on the radial coordinate is $\sim 1$ mm, which is also the average size of an hydrogel particle.}
\label{fig:flowstruct}

\end{figure}

The constant speed motion of the rising intruder begs the questions how far the flow field around the slowly moving sphere extends laterally and whether it reaches the end boundaries. To investigate this, we perform experiments with ping pong balls rising towards a free surface in a hydrogel packing in a square container illuminated by a pair of green laser sheets as shown schematically in Fig.~\ref{fig:flowstruct}. We analyze the measured flow fields around the sphere using the software PIVlab \citep{thielicke_pivlab_2014} applied to the videos recorded with the high-speed camera. The videos were pre-processed using  \textit{intensity capping} to reduce the noise from the brightest spots which arise from scattering of the laser illumination by impurities in the hydrogel suspension. We also remove the image of the ping pong ball from the region of interest for the PIV analysis using a mask.

The flow field analysis is performed using a \textit{Multipass FFT window deformation algorithm}. In particular, we used three passes with an interrogation area of 80x80, 40x40 and 20x20 pixels respectively (corresponding to approximately 18x18, 9x9 and 5x5 mm) with 50\% overlap and with the \textit{``Extreme" correlation robustness} option. This option has been shown to yield results with high signal-to-noise ratio in the presence of noise \citep{thielicke_particle_2021}. We also employ a standard approach of using different frame jumps to measure the slow velocities far away from the moving intruder~\citep{Gollin2017}.

Finally, we remove the few outliers (which is a result of out-of-plane motion or poorly lit areas) through two velocity validation algorithms, automatically implemented in PIVlab. We first filter the velocity vectors based on the magnitude of each component, requiring that it does not deviate by more than 8 standard deviations from the mean, and then we apply the \textit{local median filter} \citep{westerweel_universal_2005}. This procedure yields a Valid Detection Probability (VDP) of higher than 98\%. A typical PIV result from an experimental video is available as Supplementary Information.

\begin{figure}
\centering
{\includegraphics[scale=0.04]{fig12.pdf}}
\caption{Left: sketch of the experimental setup, with some illustrative blue trails representing the displacement of coloured hydrogel beads. Top right: superposition of a  sequence of $32$ images taken $10$ seconds apart. Bottom right: when the ping pong ball rises towards the surface, the hydrogel particles create a bulge on the surface of the packing. The black background was edited to enhance the visibility of the bulge. $C = 6.35$ g/L. }
\label{fig:flowstruct-boundary}
\end{figure}

The resulting PIV fields are averaged over $700$ frames within a domain that co-moves with the steadily rising sphere. The velocity field $u$ representing left-right motion is shown in Fig.~\ref{fig:flowstruct} (b). The quadrupolar flow structure as required by mass conservation can be clearly distinguished, and there is a weak vertical asymmetry in the profile. The velocity component parallel to gravity $v$ is shown in Fig.~\ref{fig:flowstruct} (c). The colour scheme highlights that the upward speed of the sphere is significantly higher than the return downward speed of the water in the hydrogels. This is reasonable, since the downward counterflow required by conservation of mass is spread over a wide area surrounding the sphere. To highlight the structure of the additional flow field induced by the shear between the sphere and the hydrogel sample, we constrain the colour scale in Fig.~\ref{fig:flowstruct} (d). There is a significant wake flow dragged upwards with the sphere, and there is also a large volume of hydrogel particles pushed ahead of the sphere. However, adjacent to the sphere, the hydrogels are mainly stationary, even though a no-slip boundary condition exists between the water and the sphere, hence one could expect the water-hydrogel suspension to be dragged along with the sphere, even though the sphere is not rough on the length scale of the hydrogel particles. Our measurements are accurate up to a distance of $\sim 1 $mm (approximately the size of a particle) from the spherical intruder. Combining the velocity components into $|\textbf{v}|$ permits a comparison of PIV results with tracking measurements and  the equatorial velocity component $|\textbf{v}|(r)$ with the Stokes solution, Fig.~\ref{fig:flowstruct} (e). We observe that the hydrogel fluid velocity profile has a very short decay length when compared with the long-range Stokes solution.

\subsection{Rising towards a free surface: radial displacement field }\label{subsec:radial_displacement_field}
To investigate the radial velocity profile of the hydrogel suspension during a rising sphere test, the hydrogel suspension is marked with  coloured swollen hydrogel beads, and their motion is imaged from the top of the container. The image shown in the top right panel of Fig. \ref{fig:flowstruct-boundary} is a composite constructed by superposing $32$ still images taken at $10$ second intervals during the period in which the sphere is moving towards the surface. Evidence for outward motion from the centre of the ball is clear. Upward movement of the hydrogel surface was also observed; see Fig. \ref{fig:flowstruct-boundary}, bottom right.  The flow visualisation observations in Fig. \ref{fig:flowstruct-boundary} illustrate that, despite the localization of the flow next to the intruder, the radial flow of the hydrogel packing above the sphere is substantial, and that boundary of the hydrogel fluid can deform significantly. However, as indicated with the mesh experiments, these deformation fields alone do not appear to be responsible for the creation of (sub)linear time-dependent motion of the intruder.

\begin{figure}
\centering
{\includegraphics[clip, trim=0.5cm 9cm 0.5cm 7cm,scale=0.5]{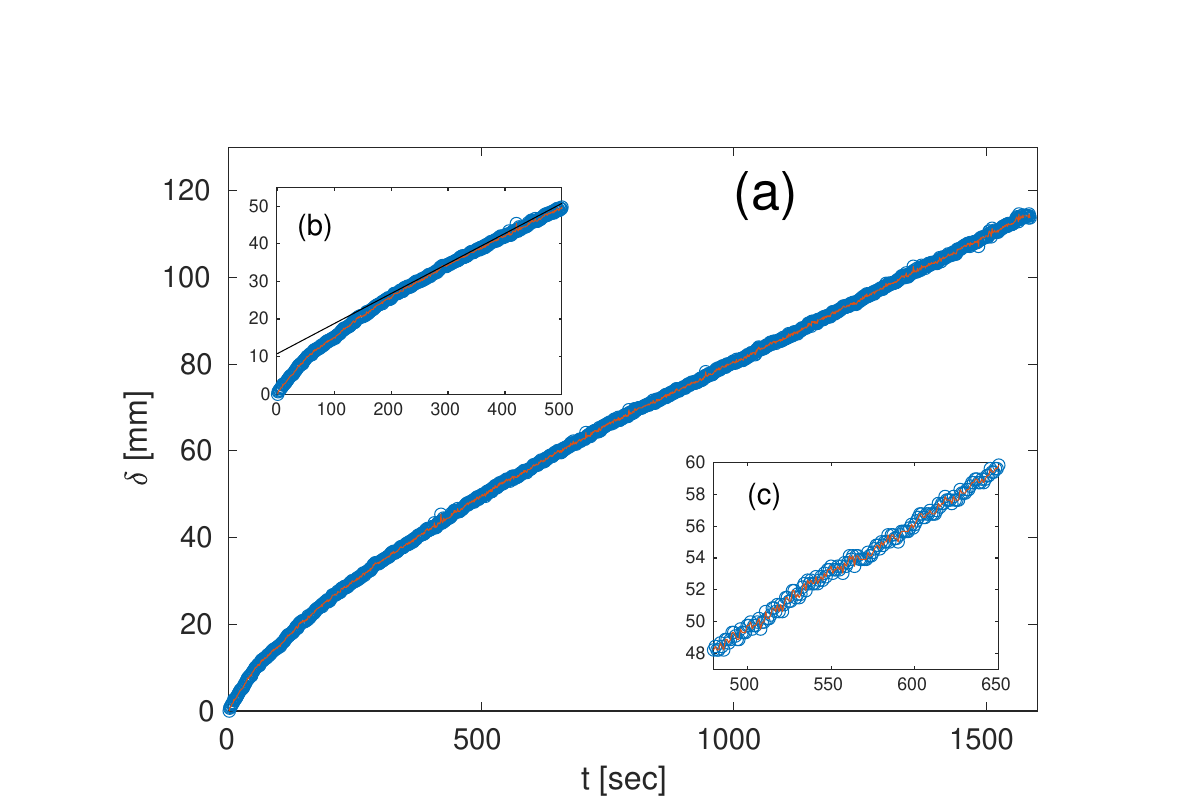}}
\caption{Motion of a ping pong ball observed using high-speed camera  and tracking software. (a) The overall behaviour is a constant velocity over after an initial transient. (b) The initial transient is a consequence of the release of the ping pong ball in the higher density lower fluid layer. After this, the motion smoothly changes into a constant velocity regime. The solid straight line is a guide to the eye. (c) In a regime of constant velocity, a close-up view highlights that the motion is smooth and there is no evidence for stick-slip behaviour.}
\label{fig:highspeed}
\end{figure}

\subsection{Smoothness of displacement}
Implicit in our arguments thus far is that the medium flows in a continuous way. In practice our sample comprises a mixture of polydisperse, deformable, porous, hydrogel particles and water which could give rise to discrete displacement steps from e.g. stick-slip or other intermittent dynamics.  We are unable to detect intermittency in the motion when we use higher spatiotemporal resolution. In Fig.~\ref{fig:highspeed} the results of  an experiment on the motion of a ping-pong ball recorded using the high speed imaging system described above show there is no evidence for discrete steps in the motion. Note that these results were obtained using the square container setup. The same qualitative behaviour was observed in separate experiments with a circular container, indicating that side-wall effects on the motion are also not significant.

\begin{figure}
\centering
{\includegraphics[clip, trim=0.5cm 5.5cm 0.40cm 6cm,width=7cm]{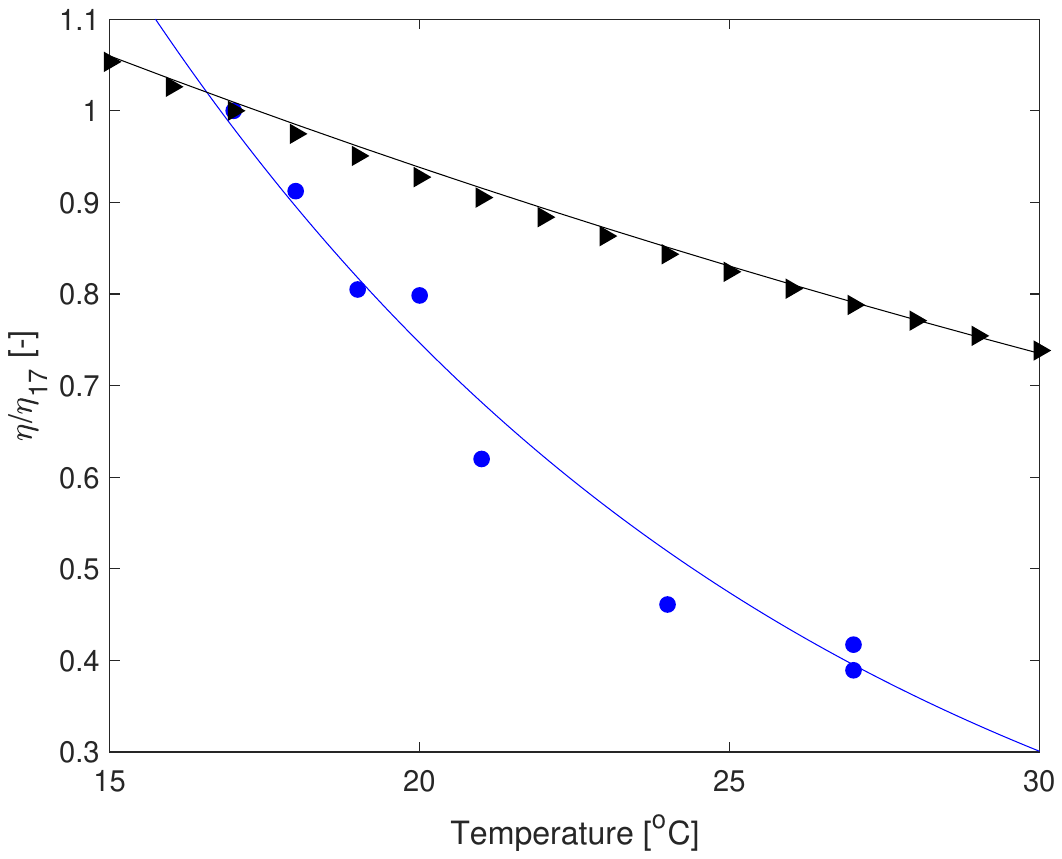}}
\caption{Effective viscosity plotted versus temperature. The hydrogel experiments (blue circles) are performed with $C = 6.35$ g/L and $\sigma_S=213$Pa. Tabulated values for water (black triangles) are provided for comparison. Data is normalized to have value unity at $17$ °C. }
\label{fig:temp}
\end{figure}

\section{Temperature and Diffusion effects}
The constant speed behaviour suggests that a water-based lubrication mechanism may play a role in the dynamics of the rising sphere. We can test this hypothesis by measuring the effective viscosity of a sample under varying environmental temperature. The sample temperature is recorded in all experiments and the buoyancy stress is kept fixed for this particular investigation. We plot in Fig.~\ref{fig:temp} the effective viscosity of a hydrogel sample versus temperature with $C= 6.35$ g/L with  a buoyancy stress of $\sigma_S=213$~Pa. The temperature range covered is $17$ to $27$ °C. Also included in Fig.\ref{fig:temp} are data points for water taken from tabulated values~\citep{huber2009new}. The data sets are normalised to be unity at $20$ °C. The fitted lines have an Arrhenius form; it can be seen that the hydrogel sample has a significantly stronger dependence on temperature than water. We  conclude that a simple water-based lubrication mechanism, for example in the observed slip between the sphere and the hydrogel sample found here and reported elsewhere~\citep{kudrollihydrogel}, cannot be the only factor that plays a role in the effective viscosity of the two-phase medium. We speculate that other temperature-dependent mechanisms contribute to the effective viscosity of the hydrogel sample. One of these could be the temperature dependence of the stiffness of the hydrogels~\citep{Dijksman2017}, which controls their poroelastic behavior~\citep{Chan2012}. However, poroelastic effects typically have a timescale that is the ratio of the stiffness and the viscosity, which for the system observed here both decrease with temperature, partially compensating this effect.

Another local mechanism that can drive the motion of an intruder is local particle diffusion in the hydrogel suspension. It is thus interesting to understand how fast the intruder moves with respect to diffusive processes in the hydrogel suspension. The intruder motion is of the order of $10^{-6}$ to $10^{-3}$ m/s. There are two types of diffusive processes involved. At the molecular level motion within the water is one, but also the hydrogel particles close to the sphere are affected by the passage of the intruder~\citep{hsiau1993shear}. More intensive Particle Tracking Velocimetry could aid in extracting the mean squared displacement $D$ of the hydrogel particles. With the intruder speed $U$ and a characteristic lengthscale $L$, one can estimate a Peclet number $Pe = UL/D$. From the available dyed tracer particle measurement as
discussed in Sec.~\ref{subsec:radial_displacement_field}, we estimate that hydrogel particles do not diffuse more than a millimeter during the course of an experiment, putting an upper bound on their mean squared displacement $D < 10^{-8}$ m$^2$/s. The diffusion constant of water is approximately $10^{-6}$ m$^2$/s, suggesting that the Peclet number should be larger than one for both the linear and sublinear scenarios. This simplified interpretation may be more subtle in practice: for example we expect $D$ to be $U$ dependent, as the diffusive motion of the hydrogels is not driven by thermal motion of the solvent molecules, but by the movement of the intruder~\citep{hsiau1993shear}, involving a self-fluidization process that is commonly observed to be present for dry granular materials ~\citep{campbell1997self, dijksman2011jamming, kamrin2024advances}. The typical lengthscale involved is also not easily identified and is dependent on the local stress~\citep{kamrin2024advances}. We thus see several avenues for future research to explore the role of microscopic mechanisms that may contribute to driving the slow flow in hydrogel suspensions.

\section{Deriving a model equation for the motion of the intruder}
\label{sec:theory}

\subsection{Derivation}

In the results presented in section \ref{end_boundary_conditions}, evidence is provided in support of a relationship between the end boundary towards which the intruder is moving and the exponent $\alpha$ of equation \eqref{eq:nonlin}. This equation describes the general power law nature of the time dependence of the trajectory of the sphere through the packing under different end boundary conditions. A distinctive property of the equation is the exponential dependence on the stress exerted by the intruder, $\sigma_S$, as shown in Fig. \ref{fig:risingnonlinear}. We now present a derivation which connects these observations, yielding the following generalized version of Eq.~\ref{eq:nonlin} as the equation of motion for the intruder 
\begin{equation}\label{eq:8nonlin_2}
    \delta = \left[ \Theta(\sigma_S, \sigma_L)t \right]^\alpha \quad ,
\end{equation}
in which both linear and sublinear dynamics  can be obtained by varying the end boundary condition. Additionally, the prefactor $\Theta$ has an exponential dependence on both the intruder  and the confinement stresses. The assumptions used in the derivation will be justified below using physical arguments based on our experimental observations. We conclude this section by providing a roadmap indicating the next steps to be taken in the development of a more rigorous theory.

We begin by considering the inertia-free motion of an intruder, as driven by its own effective weight. The driving \d{stress} is balanced by the hydrogel-induced drag 
\begin{equation}\label{eq:stress_balance}
    f(\sigma_S) = \frac{\eta_{\text{eff}}(\sigma_S, \sigma_L)\delta'}{R}
\end{equation}
Here, $\delta'$ is the derivative of the intruder displacement with respect to time and $\eta_{\text{eff}}(\sigma_S, \sigma_L)$ is the coefficient that models the drag force, which can also depend on the intruder stress, $\sigma_S$, and on the confinement stress exerted on the  packing, $\sigma_L$. Note that we use the same symbol as the effective viscosity defined in the end of section \ref{end_boundary_conditions}, but the two quantities are, in fact, only equivalent, modulo a numerical prefactor which we can safely ignore for the time being, given that we are interested in reproducing the scaling, rather than obtaining quantitative agreement. The reason why the drag coefficient can be identified as the effective viscosity will become clear later in the current section when we compare the model predictions with the experimental results, and interpret this scientific theory in terms of the physical system. Moreover, note the additional factor of $R^{-1}$, required by dimensional analysis to obtain a stress. On the left hand side of Eq.~\ref{eq:stress_balance}, the driving force $f(\sigma_S)$ is a (possibly non-linear) function of the buoyancy stress acting on the intruder, it has the dimensions of stress and encodes the bulk rheology of the hydrogel suspension. In fact, the material may have a yield stress $\sigma_Y$, which would reduce the effective driving stress, or even might possess a more complex constitutive relation of the Herschel-Bulkley type. We generically denote the effective driving stress as $\sigma_{\text{drive}} \equiv f(\sigma_S)$, and do not consider a specific type of constitutive relation. Specifically, we will show that scaling exponent of the intruder equation of motion is independent of any constitutive relationship.

The drag force coefficient  $\eta_{\text{eff}}(\sigma_S, \sigma_L)$, and in particular its dependence on the imposed stresses $\sigma_S, \sigma_L$ is now considered. We  assume an exponential dependence of the following kind
\begin{equation}\label{eq:visc_assumption}
    \eta_{\text{eff}}(\sigma_S,\sigma_L) = \eta_0 \, e^{-\frac{\sigma_S}{A_0 + k \sigma_L}}    \quad .
\end{equation}
The exponential dependence is motivated by observations on intruder and confinement stress dependence, see in particular \citep{2024creepconf}. In the above equation, $A_0$ represents a stress scale and $k$ is a prefactor. We follow with a more thorough discussion on the use of this exponential factor, after considering the connection between intruder motion $\delta$ and the confinement stress $\sigma_L$. When moving, the intruder pushes together the hydrogel particles ahead of it \textit{locally} on its trajectory towards the end boundary. This compresses the packing and increases the local confinement stress stored in the contact network. The packing may locally relax this intruder motion-induced compression, if the packing is unconstrained or the intruder motion is slow enough. Yet if the rearrangements of the hydrogel particles is reduced, for example in the case of a rigid end boundary or a lid that confines the packing, the excess confinement stress built up during the intruder movement will not relax fast enough (with respect to a time scale set by the speed of the intruder). We can formalize the proposed mechanism using the following additional assumed relation between the confinement stress and the intruder displacement
\begin{equation}\label{eq:excess_stress}
    \sigma_L (\delta)= \sigma_{L_0} + \sigma_{L_{e}}(\delta) =\sigma_{L_0} + \kappa \delta^\beta
\end{equation}
where $\beta$ is an exponent fixed by the boundary condition and $\kappa$ is a proportionality constant that plays the role of a generalized elastic modulus. Note that $\kappa$ will also depend on the boundary conditions, so that the product $\kappa \delta^\beta$ is in units of Pascals. The exponent $\beta$ can then be considered to be zero for a free end boundary: the intruder \emph{motion} does not result in an additional stress build-up. A rigid boundary ahead of the intruder induces stronger compaction of the packing, hence $\beta>0$.  See Fig. \ref{fig:mechanism} for a pictorial description of the different cases. Inserting equation \eqref{eq:excess_stress} into our ansatz \eqref{eq:visc_assumption}, and assuming that the additional excess confinement $\sigma_{L_e}$ is small with respect to $\sigma_{L_0}$, hence that $\kappa$ is small, we can approximate the effective viscosity as
\begin{equation}\label{eq:eff_visc_approx2}
    \eta_{\text{eff}}\left(\sigma_S,\sigma_L(\delta)\right) \approx \tilde{\eta}_{\text{eff}} \left( 1 + \frac{k \sigma_S}{(A_0 + k\sigma_{L_0})^2} \kappa \delta^\beta \right) \quad .
\end{equation}
In this expression, $\tilde{\eta}_{\text{eff}}$ includes both the prefactor $\eta_0$ and the order zero exponential factor $\exp{\left(-\frac{\sigma_S}{A_0 + k \sigma_{L_0}}\right)}$. We insert the expression for the displacement-dependent effective viscosity into the force balance equation \eqref{eq:stress_balance} and obtain solutions for the trajectory of the intruder. In particular, the differential equation has the form
\begin{equation}
     \Sigma = \left(1 + \Gamma \delta(t)^\beta\right) \delta'(t)
\end{equation}
where we make explicit the time dependence of the intruder displacement, and where we introduce the constants $\Sigma$ and $\Gamma$
\begin{equation}
    \Sigma = \frac{\sigma_{\text{drive}} R}{\tilde{\eta}_{\text{eff}}} \quad , \quad \Gamma = \kappa\frac{k \sigma_S}{(A_0 + k \sigma_{L_0})^2} \quad .
\end{equation}
The differential equation is separable and can be solved by the following implicit equation
\begin{equation}\label{eq:implicit_sol}
    \Sigma  t = \delta(t) + \Gamma\frac{\delta(t)^{\beta+1}}{\beta+1} \quad ,
\end{equation}
which is not invertible for any arbitrary exponent $\beta$. We can anyway obtain that at large time $t > t_0$ the displacement scales as $\delta \sim t^{\frac{1}{\beta +1}}$ when taking into account the additional constraints that the observed displacement is a monotonously increasing function, and that $\beta \ge 0$. This scaling corresponds to the desired behaviour, equation \eqref{eq:8nonlin_2}. As desired, the exponent $\alpha$ is completely determined by the \textit{rigidity} $\beta$ of the boundary condition according to the relation $\alpha = 1 / (\beta + 1)$.\\ 
The timescale $t_0$ is set by the parameters of the problem and represents the time at which the initial transient (which scales linearly crosses over to the late time power-law scaling. The time $t_0$ can be obtained by comparing the linear and non-linear contribution in equation \eqref{eq:implicit_sol}, which yields $t_0 = \frac{1}{\Sigma}\left( \frac{\beta + 1}{\Gamma}\right)^{\frac{1}{\beta}}$. Note, though, that the initial transient in the experimental measure of $\delta(t)$ is strongly affected by the intruder release mechanism, and is therefore unlikely to be well captured by our simplified model.

Equation \eqref{eq:implicit_sol} can be readily solved, instead, for $\beta = 0$ (which we associate to the free end boundary case), yielding linear motion of the intruder
\begin{equation}
    \delta(t) = \frac{\Sigma}{\Gamma + 1} t
\end{equation}
with a velocity proportional to the inverse effective viscosity $\tilde{\eta}_{\text{eff}}$ (which depends exponentially on the intruder stress). When $\beta >0$, the effective viscosity $\eta_{\text{eff}}$ increases with the intruder displacement, see equation \eqref{eq:eff_visc_approx2}, and we obtain sublinear motion. Note, in fact, that an implication of \eqref{eq:stress_balance} is that for a constant imposed stress $\sigma_S$ we must have $\eta_{\text{eff}}(t)\sim \delta'^{-1}(t)$ at all times.

\begin{figure}
\centering
{\includegraphics[scale=0.6]{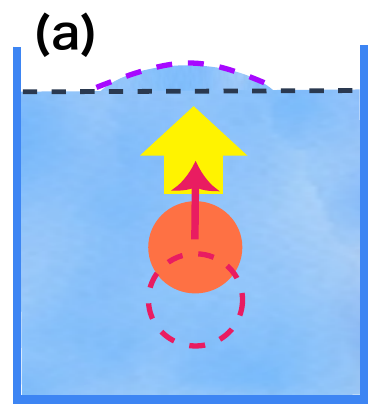}}
{\includegraphics[scale=0.6]{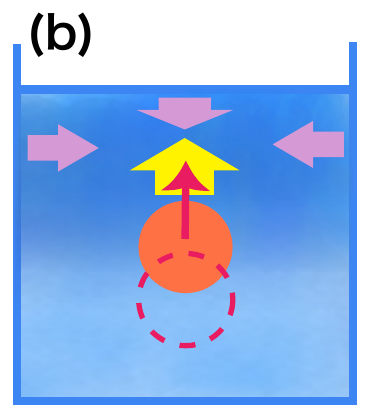}}
{\includegraphics[scale=0.6]{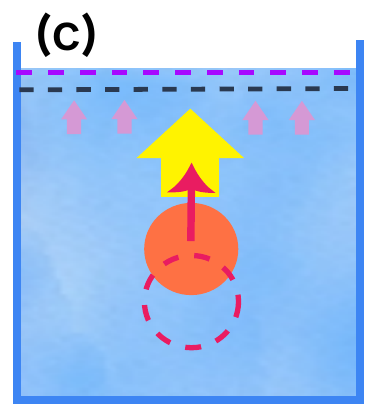}}
\caption{A schematic of the response of the hydrogel packing to a rising intruder under different boundary conditions. (a) As the intruder rises towards a free surface, the internal stress of the packing relaxes vertically by expanding in the direction of motion, creating a bulge on the surface of the packing (the purple dashed line represents the deformed surface of the material). (b) A rigid end boundary prevents relaxation of the stress. Instead, it diffuses (isotropically) through the contact network. This produces a slight increase in the local density (represented in the picture by a darker shade of blue) which consequently leads to an increase in the apparent viscosity. (c) A flexible, light metallic mesh is placed at the surface. The mesh is flexible but has a finite bending stiffness. It is light and the relaxation of the stress causes the mesh to rise uniformly rather than bend as would be necessary to create a bulge. Note the volume of material displace is exaggerated  in panel (c) to clarify the effect of the mesh.}
\label{fig:mechanism}
\end{figure}

The case $\beta = 1$,  corresponds to a fully elastic local response of the packing ahead of the intruder as in the case of the rigid end boundary at the bottom of a container. In this situation, the full solution is
\begin{equation}
    \delta(t) = \frac{1}{\Gamma} \left( \sqrt{2 \Sigma \Gamma t + 1} - 1 \right)
\end{equation}
and we retrieve the square root scaling observed in experiments of sinking intruders as in \citep{2022creepcontrol}.

\subsection{Interpretation}
The interpretation for the \textit{linear-to-sublinear} transition of the intruder trajectory discussed here relies on a number of assumptions that are in accord with observations. The two main assumptions on which the derivation rests are: (i) the functional form of the drag force coefficient (termed effective viscosity), equation \eqref{eq:visc_assumption}, and in particular its exponential dependence on the intruder stress and on the confinement stress; (ii) the \textit{local compression} mechanism that we envision for the increase of the viscosity during the intruder motion. In addition, the following hypotheses are assumed to hold: (iii) the motion  of the intruder is inertia-free; (iv) the excess confinement stress $\sigma_{L_e}$ is small with respect to the other stress scales involved, allowing for the expansion \eqref{eq:eff_visc_approx2} of the effective viscosity.

For (i), the choice of an exponential dependence of the effective viscosity $\eta_{\text{eff}}$ on $\sigma_S$ and on $\sigma_L$ is motivated by the observation of a functional relation of the same type for the prefactor $\Theta$ of the intruder time trajectory. As mentioned in section \ref{end_boundary_conditions}, an exponential dependence on $\sigma_S$ was established in previous studies \citep{2022creepcontrol, 2024creepconf}. There, the intruder was sinking towards the rigid bottom of a container, and the displacement was found to follow the relation \eqref{eq:8nonlin_2} with $\alpha = 1/2$. In the current study we confirm this relation, observing $\Theta \propto \exp(\sigma_S)$ over a wide parameter range in the experiments: various hydrogel concentrations $C$, different intruder sizes and multiple boundary conditions (free surface and lid). The prefactor $\Theta$, hence, plays the role of a generalized velocity (the actual speed $U$ in the linear case, the diffusion coefficient $D$ for the square root case) allowing a connection to be established with the generalized drag force coefficient $\Theta \sim \eta_{\text{eff}}^{-1}$, which we hence equivalently call \textit{effective viscosity}, since for $\alpha = 1$ it simply reduces to that. The effect of confinement stress $\sigma_L$ was also studied in \citep{2024creepconf} for a sinking intruder. It was found that the confinement pressure sets the stress scale for the creep rate and that it is possible to collapse all the experimental results using the relation
\begin{equation}\label{eq:full_stress_D_dep}
    D = D_0 e^{\frac{\sigma_S}{A_0 + k \sigma_L}}
\end{equation}
This supports the use of this functional form as the ansatz for the effective viscosity stress dependence.

The second main assumption (ii) concerns the response of the packing when the intruder moves. Our hypothesis, illustrated schematically in Fig. \ref{fig:mechanism}, is that the intruder acts as a moving rigid boundary which compresses the hydrogel particles in front of it. The rigidity of the end boundary does not allow the particles to move radially as instead observed in the case of a free boundary (cfr section \ref{subsec:radial_displacement_field}), so the internal stresses cannot relax as easily or as quickly, and locally retains a surplus stress
\begin{equation}\label{eq:excess_stress2}
    \sigma_{L_e} = \kappa \delta^\beta
\end{equation}
This excess stress originates from the elastic compression of hydrogel particles which, confined between the intruder and the (rigid) end boundary, have a limited capacity for rearrangement. Equation \eqref{eq:excess_stress2} encodes this elastic response of the hydrogel, which will be localized in part of the packing. The prefactor $\kappa$ includes information on the volume over which the elastic compression is spread, and depends on the relative geometries of the intruder and of the container. To get an idea of why this is the case, consider the (elastic) excess stress that builds up after a change in volume, $\sigma_{L_e} \propto K\frac{\Delta \rho}{\rho} = - K \frac{\Delta V}{V}$, with $K$ the bulk modulus. When the intruder moves, it sweeps a volume proportional to $\pi R^2 \delta$. According to our hypothesis of local compression, the change in volume $\Delta V$ will be proportional exactly to the volume swept by the ping pong ball (to some power $\beta$), and the excess stress will then read
\begin{equation}
    \sigma_{L_e} \propto K \left( \varsigma \tilde{h}^{-1} \delta\right)^\beta
\end{equation}
with $\varsigma$ the ratio between the section of the intruder, $\pi R^2$, and the section of the container, and with $\tilde{h}$ the height of the portion of hydrogel material in which the elastic compression is distributed. We can hence recognise the constant $\kappa$ as the product of the bulk modulus with a geometry dependent coefficient, $\kappa \propto K (\varsigma \tilde h^{-1})^\beta$. The exponent $\beta$, instead, controls the rigidity of the end boundary: for $\beta = 1$ there is a perfectly rigid response of the boundary; for $\beta = 0$ no excess stress is accumulated, corresponding to a free end surface, which allows the individual hydrogel particles to relax all stresses arising from the motion of the intruder; in all the other intermediate cases, $0 < \beta<1$, the end boundary partially constrains the packing (e.g. a light lid can tilt or move), resulting in a partial compression of the hydrogel grains. Relation \eqref{eq:excess_stress2}, thus, provides an effective description of the coupling between the boundary conditions and the intruder motion. We note that in the limit of very small intruders, $\varsigma \to 0$, the compression effect may disappear as well, yielding constant velocity motion even for intruders sinking towards a rigid boundary, as discussed via Fig.~\ref{fig:smallspheres}.

Finally, we address the last two additional assumptions.  (iii) The inertia-free balance equation is justified since the effective $Re \ll 1$. (iv) The expansion of the effective viscosity performed in equation \eqref{eq:eff_visc_approx2} is also justified since the scale of both the intruder stress and the pre-existing confinement stress $\sigma_{L_0}$ is much larger than the additional stress coming from intruder-induced compression of the packing.

We conclude this section by commenting on the independence of the temporal exponent $\alpha$ from initial compaction. This result might be surprising at a first glance, for a material where history-dependent effects are known. According to our theory, though, it is only the global constraints that determine the scaling of the compression with time (which in turn determines the functional form of the equation of motion of the intruder). Preparation-dependent factors such as initial compression only affect the numerical prefactors. In particular, the initial compaction $\sigma_{L_0}$ is encoded in the parameter $\Gamma$, which exclusively affects the slope of the intruder trajectory, and not its scaling. Our simple theory, hence, offers a way to disentangle effects due to global geometrical constraints from those coming from preparation-dependent details.

\section{Discussion \& Outlook}

Further work is required to substantiate the proposed mechanisms. A crucial step would be to explore the dependence of the confinement stress for other boundary conditions: equation \eqref{eq:full_stress_D_dep} has, thus far, only been validated for a fully rigid boundary condition. Moreover, the functional form of the effective viscosity suggests the existence of an exponential constitutive relation between stresses and deformation rates for the continuum description of hydrogel packings. Experimental observations of  an exponential relation also in the bulk rheology of the material would support our claim.

Another important test of this interpretation would be the validation of equation \eqref{eq:excess_stress2}. The degree of local compaction in the hydrogels could be measured around the intruder, and, hence, spatiotemporal variations of the density could be measured.   This is beyond the capabilities of our current experiments but it could be done using  facilities with greatly improved spatial resolution.
Measurements of bulk rheology could also provide insights into relaxation dynamics of the packing by probing, e.g., changes in the lid position with time. This would enable an understanding of the relaxation of the intruder related excess stresses, and test our hypothesis that the surplus stress $\sigma_{L_e}$ does not relax on the  time scale of the experiment. To pinpoint the microscopic origin of the long range effect of the boundaries, it may be insightful to extend the current studies with those on frictional hydrogel particles~\citep{workamp2019contact}. Finally, the role of non-local flow models~\citep{kamrin2024advances} on the boundary sensitivity should also be clarified, as such models have a similar tendency to show long-range effects.

\section{Conclusions}
We systematically tested the influence of the end boundary of a packing of soft spheres on the drag force exerted on an intruder moving inside the packing. We exposed the packings to a constant intruder stress  using hollow plastic spheres and observed their motion in both sinking and rising. We found that spheres travel upwards with $\delta \propto t$ in cases with a free surface on the hydrogel packing. We contrast the constant velocity dynamics with creep flow found earlier to confirm a dependence of the flow behavior on the boundary conditions. We find that when a lid is placed on the surface of the material, a rising sphere exhibits $\delta = (\Theta t)^{\alpha}$, with $\alpha \neq 1$. We show that the qualitative difference between the constant velocity and material creep motion is accompanied by clear evidence of flow at the free surface. The velocities or creep rates achieved are found to depend exponentially on the buoyancy stress i.e. $\propto \exp{(\sigma_S)}$, consistent with earlier work. We confirm that the exponential stress dependence is not a solvent viscosity effect by performing a temperature tests. We use flow field visualization to indicate the local nature of the flow around the intruder. We hypothesize a quantitative connection between the change in boundary conditions, the exponential packing stress dependence and the time-dependent flow behavior, stimulating theoretical progress while focusing future work on the as yet unexplained but consistently observed exponential stress dependence of the flow behavior of hydrogel packings.

 \section*{Acknowledgments}
The authors are grateful to Keith Long for constructing the apparatus used in the majority of the experiments. We also thank several undergraduate students at the University of Amsterdam for their preparation work on some rising sphere experiments. Finally, we are grateful to Dominic Vella for his suggestion of using a mesh as the upper boundary and to the Leverhulme Trust who supports TM's research through an Emeritus Fellowship EM-2024-014/4.

\medskip

\noindent Declaration of Interests. The authors report no conflict of interest.

\bibliography{ppong}

\bibliographystyle{jfm}

\end{document}